\documentclass[aps,prb,reprint,twocolumn,superscriptaddress,floatfix,nofootinbib]{revtex4-2}
\usepackage{epsfig,amsmath,amssymb,color,comment,physics, dsfont, bbold,amsfonts}
\usepackage[makeroom]{cancel}
\usepackage[caption=false]{subfig}
\usepackage{mathrsfs}
\usepackage[countmax]{subfloat}
\usepackage[normalem]{ulem}
\usepackage[english]{babel}
\usepackage{dsfont}
\usepackage{float}
\usepackage{color}
\usepackage[dvipsnames]{xcolor}
\usepackage{tikz}

\usepackage[bookmarks=true,colorlinks,linkcolor=blue,urlcolor=NavyBlue,citecolor=RoyalBlue]{hyperref}

\newcommand{\defaulttensorsize}{10pt}
\newcommand{\tensorsize}{\defaulttensorsize}

\usetikzlibrary{patterns,patterns.meta}
\usetikzlibrary{shapes.geometric}
\usetikzlibrary{shapes.misc}
\tikzstyle{tensor}=[draw, inner sep=0, outer sep=0, minimum size=\tensorsize]
\tikzstyle{notensor}=[inner sep=0, outer xsep=2pt, outer ysep=0, minimum size=\tensorsize]
\tikzstyle{atensor}=[tensor]
\tikzstyle{ctensor}=[tensor, circle]
\tikzstyle{dtensor}=[tensor, diamond]
\tikzstyle{wtensor}=[tensor]
\tikzstyle{ltensor}=[tensor, rounded rectangle, rounded rectangle left arc=none]
\tikzstyle{rtensor}=[tensor, rounded rectangle, rounded rectangle right arc=none]
\tikzstyle{etensor}=[tensor, minimum height=(1cm/\defaulttensorsize*0.5*2+1)*\tensorsize]
\tikzstyle{widetensor}[2]=[tensor, minimum width=(1cm/\defaulttensorsize*0.75*(#1-1)+1)*\tensorsize]
\tikzstyle{tensornetwork}=[baseline=-0.25em, xscale=0.75, yscale=0.5,
                           every node/.style={inner sep=0, outer xsep=2pt, outer ysep=5pt, text depth=0pt},
                           execute at end picture={\path (current bounding box.south west) +(-2pt, 0) (current bounding box.north east) +(2pt, 0);}]

\tikzset{round left paren/.style={ncbar=0.5cm,out=120,in=-120}}
\tikzset{round right paren/.style={ncbar=0.5cm,out=60,in=-60}}

\graphicspath{}
\usetikzlibrary{quantikz}

\begin{document}

\title{Spectral signatures of nonstabilizerness and criticality in infinite matrix product states}

\author{Andrew Hallam}
\thanks{These authors contributed equally.}
\affiliation{School of Physics and Astronomy, University of Leeds, Leeds LS2 9JT, United Kingdom}

\author{Ryan Smith}
\thanks{These authors contributed equally.}
\affiliation{School of Physics and Astronomy, University of Leeds, Leeds LS2 9JT, United Kingdom}

\author{Zlatko Papi\'c}
\affiliation{School of Physics and Astronomy, University of Leeds, Leeds LS2 9JT, United Kingdom}

\date{\today}
\begin{abstract}
While nonstabilizerness (``magic”) is a key resource for universal quantum computation, its behavior in many-body quantum systems, especially near criticality, remains poorly understood. We develop a spectral transfer-matrix framework for the stabilizer Rényi entropy (SRE) in infinite matrix product states, showing that its spectrum contains universal subleading information. In particular, we identify an SRE correlation length -- distinct from the standard correlation length -- which diverges at continuous phase transitions and governs the spatial response of the SRE to local perturbations. We derive exact SRE expressions for the bond dimension $\chi=2$ MPS ``skeleton” of the cluster–Ising model, and we numerically probe its universal scaling along the $\mathbb{Z}_2$ critical lines in the phase diagram. These results demonstrate that nonstabilizerness captures signatures of criticality and local perturbations, providing a new lens on the interplay between computational resources and emergent phenomena in quantum many-body systems. 
\end{abstract}
\maketitle

\section{Introduction}

The implementation of a universal quantum gate set is a major challenge in large-scale, fault-tolerant quantum computation~\cite{Nielsen_Chuang_2010}. While only Clifford operations are typically feasible, forming a subset that admits efficient classical simulation~\cite{gottesman1998,Eastin2009}, universal computation can be achieved by injecting nonstabilizer or ``magic'' states into circuits~\cite{knill2004,Bravyi2005,Campbell2010}. This raises the practical question of how such special states can be generated and manipulated. Although significant progress has been made in understanding nonstabilizerness in few-qubit systems~\cite{Bravyi2005,Howard2017}, its role in many-qubit Hamiltonian and circuit settings remains an active area of investigation~\cite{ShiyuTgate2020,goto2021chaos,Liu2022,Leone2022,Haug2023MPS,Tarabunga2023, Turkeshi2023,Bejan2024circuits,niroula2024phase,Tarabunga2024RK,Haug2025probingquantum}.

At this stage, many basic questions surrounding nonstabilizerness remain open, such as whether it can play a similar role to entanglement in characterizing the universal properties of many-body systems. Indeed, quantum entanglement is now central to the understanding of exotic phases of matter~\cite{Kitaev2006,Levin2006,Li2008,Pollmann2012}, and the dynamics of interacting quantum systems~\cite{Bardarson2012,Serbyn2013,lukin2019} (see Ref.~\cite{Laflorencie2016} for a review). Furthermore, low-energy eigenstates of quantum Hamiltonians typically possess a limited amount of entanglement, making them amenable to variational ansatze known as matrix product states (MPS)~\cite{SCHOLLWOCK201196,CiracRMP} and powerful numerical algorithms such as the density-matrix renormalization group (DMRG)~\cite{WhiteDMRG}. 

Compared to entanglement, quantifying the nonstabilizerness of many-body wave functions is generally far more costly. In this work we focus on the stabilizer Rényi entropy (SRE)~\cite{Leone2022,Oliviero2022}, a recently proposed measure for many-qubit wave functions (the related local measures include the robustness of magic~\cite{Heinrich2019robustnessofmagic,Howard2017,Sarkar_2020} and mana entropies for qudit systems~\cite{tarabunga2024critical,frau2024nonstabilizerness}). While the SRE is still exponentially hard to evaluate, it admits Monte Carlo approximations~\cite{Tarabunga2023,lami2023quantum,Liu2025Ising}, and for MPS with sufficiently low bond dimension it can even be computed in closed form~\cite{Haug2023MPS,Banuls2024}. These tools have enabled recent insights into  nonstabilizerness in a plethora of many-body settings, e.g., critical systems~\cite{Tarabunga2023,tarabunga2024critical,Ding2025singularites,viscardi2025interplayentanglementstructuresstabilizer,hoshino2025stabilizerrenyientropyconformal,moca2025nonstabilizernessdiagnosticcriticalityexceptional,catalano2025SREtopphases}, maximally scrambling models~\cite{bera2025nonstabilizernesssachdevyekitaevmodel,zhang2025stabilizerrenyientropytransition,jasser2025stabilizerentropyentanglementcomplexity,Russomanno2025SYK,Turkeshi2025randomqc}, different types of quantum dynamics~\cite{falcao2025magicdynamicsmanybodylocalized,hernándezyanes2025nonstabilizernessquantumenhancedmetrologicalprotocols,odavić2025stabilizerentropynonintegrablequantum,tirrito2025anticoncentrationnonstabilizernessspreadingergodic,López_2024XXZ,moca2025quantumwalk} including operator dynamics~\cite{Dowling2025Heisenberg,dowling2025bridgingentanglementmagicresources}, and systems of identical particles~\cite{collura2025quantummagicfermionicgaussian,sierant2025fermionicmagicresourcesquantum,sarkis2025magichybridbosonfermionsystems,wang2025magictransitionmonitoredfree,crew2025magicentropyhybridspinboson}. 

\begin{figure}
    \centering
    \includegraphics[width=0.99\linewidth]{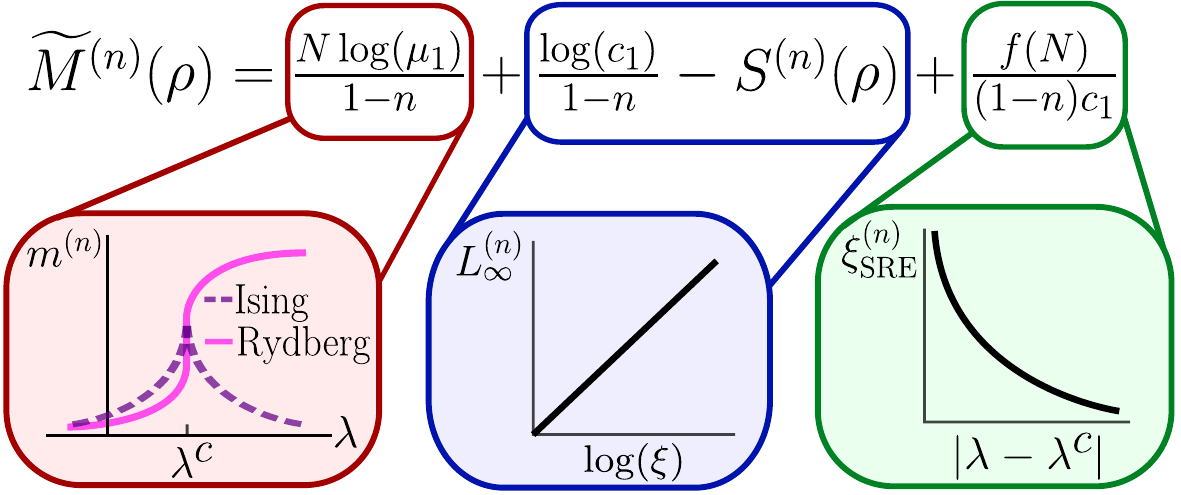}
    \caption{Our central result, Eq.~(\ref{Eq:Magic Universal}), for the mixed state SRE, $\widetilde{M}^{(n)}$, of order $n$. The density matrix $\rho$ describes an $N$-qubit subsystem of an infinite MPS  state. $\widetilde{M}^{(n)}$ splits into three boxed terms. The red box is an extensive term $\propto N m^{(n)}$ due to the dominant eigenvalue $\mu_1$ of the replica transfer matrix. This term is non-universal, as illustrated by its different behavior for the Ising-spin and Rydberg-atom realizations of the same $\mathbb{Z}_2$ critical point upon varying $\lambda$. The blue box represents correlations between the subsystem and its boundary, determined by the dominant eigenvector of the replica transfer matrix ($c_1$ term) and the R\'enyi entropy $S^{(n)}$. This term defines the mutual SRE,  $L_\infty^{(n)}$, of two adjacent semi-infinite subsystems, which diverges logarithmically with the correlation length $\xi$. Finally, in the green box, $f(N)$ represents the subleading, exponentially-decaying contribution to the SRE. This defines the SRE correlation length, $\xi_\mathrm{SRE}^{(n)}$, which exhibits a power-law divergence near criticality.
    } 
    \label{fig:equation contributions}
\end{figure}

Despite much progress, a key question remains: what does the SRE truly reveal about a wave function? The ground states of many spin-chain models exhibit varying amounts of nonstabilizerness but rarely appear to saturate the SRE bound, even at criticality~\cite{Tarabunga2023}. The lack of understanding of this behavior underscores the need for analytically-tractable models where the SRE enhancement, beyond single qubits, can be rigorously established~\cite{smith2024Rydbergatoms}. Moreover, by analogy with entanglement in condensed matter~\cite{Calabrese2004}, it would be desirable to identify universal properties encoded in nonstabilizerness. Numerical studies of certain models have indeed reported universal SRE scaling~\cite{Tarabunga2023,cao2025gravitationalbackreactionmagical,Passarelli2024permutationallyinvariant,frau2024nonstabilizerness}, supported in some cases by conformal field theory (CFT)~\cite{hoshino2025stabilizerrenyientropyconformal,hoshino2025stabilizerrenyientropyencodes} and exact calculations for non-interacting systems~\cite{rajabpour2025stabilizershannonrenyiequivalenceexact}. However, the general reliability of the SRE as a diagnostic of criticality remains unclear. For example, while the critical Ising model displays non-analytic features in the SRE~\cite{Tarabunga2023,dóra2024momentumspacemagictransverse,Leone2022Ising}, a Rydberg-atom model realizing the same $\mathbb{Z}_2$ critical point shows smooth behavior of the SRE~\cite{Falcao2025U(1),smith2024Rydbergatoms}. These contrasting observations raise doubts about whether nonstabilizerness alone can serve as a robust indicator of phase transitions.

In this work, we develop a spectral framework for nonstabilizerness in infinite matrix product states (iMPS) using the eigenspectrum of their SRE replica transfer matrices~\cite{Haug2023MPS,liu2025SRETIMPS}. We show that the SRE of a subsystem embedded in an infinite chain generally decomposes into three contributions (Fig.~\ref{fig:equation contributions}): (i) an extensive, model-dependent term capturing global nonstabilizerness; (ii) a boundary term corresponding to the mutual SRE between two semi-infinite subsystems; and (iii) subleading terms that lead to exponentially-decaying SRE correlations. This decomposition allows us to define an SRE correlation length that diverges at continuous phase transitions and governs the spatial response to local perturbations. For the exactly solvable bond dimension $\chi=2$ MPS skeleton of the cluster–Ising model~\cite{Wolf2006Skeleton,Smith2022Ctopologicalphase}, these quantities can be obtained analytically, offering microscopic insight into the behavior of nonstabilizerness. Applying the same formalism to the full cluster–Ising model, we map the SRE across its phase diagram and probe universal aspects of SRE scaling along the $\mathbb{Z}_2$ critical line.

The remainder of the paper is organized as follows. Section~\ref{sec:overview} provides an overview of the SRE and its evaluation for iMPS. In Sec.~\ref{sec:magic_subsystems}, we derive the three-part decomposition of the SRE highlighted in Fig.~\ref{fig:equation contributions}, and we introduce the SRE correlation length. Section~\ref{sec:quantum_circuits} demonstrates that this length scale governs the response to local perturbations. We then illustrate this spectral framework using the analytically tractable MPS skeleton of the cluster–Ising model in Sec.~\ref{sec:MPS_skeleton}, while the full phase diagram of this model and the universal SRE scaling along its critical lines are discussed in Sec.~\ref{sec:fullclusterIsing}. Finally, Sec.~\ref{sec:conclusions} summarizes our findings, while the Appendices contain technical details, extensions to non-adjacent subsystems, comparisons with alternative nonstabilizerness measures, exact diagonalization benchmarks on finite systems, and the analysis of a $\chi=4$ skeleton exhibiting a richer spectrum of correlations. 

\section{Overview of infinite matrix product states and stabilizer R\'enyi entropies}\label{sec:overview}

In this section, we introduce the concept of stabilizer Rényi entropies (SREs), a measure of nonstabilizerness that can be efficiently computed using a replica trick. We also review the general theory of correlations in iMPS and how they can be computed from the associated transfer matrix. This section does not contain new results, but sets the notation and background for our main results in the following Secs.~\ref{sec:magic_subsystems}-\ref{sec:quantum_circuits}.

\subsection{Scaling of correlations in iMPS} \label{subsec:SRE corrections}

A translationally-invariant iMPS state $|\psi(A)\rangle$ is defined on an infinite one-dimensional chain with a $d$-dimensional local Hilbert space,
\begin{eqnarray}\label{eq:iMPS}
\ket{\psi(A)} = \sum_{\{\sigma_{j}\}} \mathrm{tr}\left(\cdots A^{\sigma_{j-1}} A^{\sigma_{j}} A^{\sigma_{j+1}} \cdots\right) \ket{ \{\sigma_j\} },
\end{eqnarray}
where $A^{\sigma_j}$ is a set of $d$ matrices with bond dimension $\chi$. Below we will mainly be interested in the case $d=2$, although the derivations hold for any $d$. Furthermore, we assume one-site translation invariance for simplicity; however, the generalization to $k$-site translationally invariant systems is straightforward and merely requires enlarging the unit cell accordingly.

An important object associated with an iMPS is its $\chi^2\times \chi^2$ \emph{transfer matrix} $E$~\cite{CiracRMP}, with the following diagrammatic notation and spectral decomposition:
\begin{equation}\label{eq:TMspectral}
    E = \sum_{\sigma} A^{\sigma} \otimes (A^{\sigma})^{*} \equiv 
    \begin{tikzpicture}[tensornetwork, scale=1.2]
    \renewcommand{\tensorsize}{16pt}
        \node[atensor,label={center, text depth=}:\(A^{*}\)] (A1) at (0, 1) {};
        \node[atensor,label={center, text depth=}:\(A\)] (A1conj) at (0, -1) {};
        \draw (A1) -- (A1conj);
        \draw (A1.west) -- +(-0.5, 0);
        \draw (A1.east) -- +(0.5, 0);
        \draw (A1conj.west) -- +(-0.5, 0);
        \draw (A1conj.east) -- +(0.5, 0);
    \end{tikzpicture}
     = \sum_{i=1}^{\chi^2} \lambda_{i}|R_{i})(L_{i}|,
    \end{equation}
where the eigenvalues $\lambda_{i}$ are ordered in descending magnitude, and $(L_{i}|$ and $|R_{i})$ denote the corresponding left and right eigenvectors. For the tensor $A^{\sigma}_{i,j}$ to represent a normalized state, the dominant eigenvalue $\lambda_1 = 1$, which we assume to be unique. The associated left and right eigenvectors correspond to the infinite contraction of $E$ from the left and right, respectively, obeying the normalization condition $(L_{1}|R_{1}) = 1$.

Every (injective) MPS satisfies an entanglement area law with  exponentially decaying two-point correlation functions:
\begin{equation}\label{eq:arealaw}
 S_E(L_A)\leq\log\chi, \quad \langle \hat{O}_n \hat{O}_{n+m} \rangle_c \sim e^{-m/\xi},   
\end{equation}
where $S_E(L_A)$ is the von Neumann entropy of a subsystem of size $L_A$, $\langle \ldots \rangle_c$ denotes the connected correlator and $\xi$ is the correlation length~\cite{CiracRMP}. The correlator in Eq.~\eqref{eq:arealaw}  is evaluated by $(m{-}1)$ applications of the MPS transfer matrix, hence its decay is  determined by the spectrum of this matrix. In particular, the slowest decay is governed by the second-largest eigenvalue in Eq.~\eqref{eq:TMspectral},
\begin{equation}\label{Eq:correlation length}
    \xi=\frac{-1}{\text{log}|\lambda_2|} 
\end{equation}
which defines the MPS correlation length.

In contrast to Eq.~\eqref{eq:arealaw}, quantum critical points in one dimension generally exhibit logarithmically divergent entanglement entropy and algebraically-decaying correlation functions. For critical points described by a $(1+1)$-dimensional CFT~\cite{Calabrese2004}, those take the form:
\begin{equation}\label{eq:criticalentropy}
S_E(L_A) \sim (c/6) \log{L_A}, \quad  \langle \hat{O}_n \hat{O}_{n+m} \rangle_c \sim |m|^{-\alpha},
\end{equation}
where $c$ is the central charge and $\alpha$ is the corresponding operator scaling dimension. Eq.~(\ref{eq:criticalentropy}) cannot be exactly captured by a finite-bond-dimension iMPS; instead, the finite bond dimension acts as a \emph{relevant perturbation} to the critical ground state, introducing an effective gap that enforces an entanglement area law and exponentially decaying correlation functions, as in Eq.~(\ref{eq:arealaw}). As the bond dimension $\chi$ increases, this perturbation diminishes, and the iMPS provides an increasingly accurate approximation of the true critical ground state. 

At criticality, the correlation length $\xi$ and, more generally, all quantities of the form $(\log|\lambda_i| - \log|\lambda_j|)^{-1}$ diverge, as they represent inverse length scales of the system. Such length scales naturally correspond to the system size in finite-size scaling analyses used to extract critical properties from finite systems. In the MPS context, however, the control parameter is the bond dimension rather than the system size, and the corresponding framework is known as \emph{finite entanglement scaling}~\cite{Tagliacozzo2008,Pollmann2008}. For a critical iMPS, the entanglement entropy $S_E(\chi)$ and correlation length $\xi(\chi)$ at finite bond dimension are related via $S_E(\chi) = (c/6) \log{\xi(\chi)}$, which mirrors the usual relationship between entanglement entropy and system size, Eq.~(\ref{eq:criticalentropy}), with $\xi$ being an effective system size.

\subsection{Stabilizer R\'enyi entropies for iMPS}\label{sec:SRE}

A convenient monotone of nonstabilizerness is the SRE of order $n$~\cite{Leone2022}. For a pure state $\ket{\psi}$ of a system containing $L$ qubits, the SRE is defined as :
\begin{equation}\label{eq:SRE}
M^{(n)}(\ket{\psi})=(1-n)^{-1}\log{\sum_{P\in\mathcal{P}_L}\frac{\bra{\psi}P\ket{\psi}^{2n}}{2^L}},
\end{equation}
where  $\mathcal{P}_L$ denotes the set of all $L$-strings of Pauli matrices $\{\sigma^\alpha\}=\{\mathbb{I},\sigma^x,\sigma^y,\sigma^z\}$. 
The SRE is zero iff $\ket{\psi}$ is a stabilizer state; it is invariant under Clifford unitaries and additive under tensor product~\cite{Leone2022}. Since our focus is on infinite systems, we also define the SRE density $m^{(2)}=M^{(2)}/L$ as an intensive measure of nonstabilizerness. Unlike other monotones, the SRE is not a strong monotone but it is more straightforward to calculate for many-body systems \cite{Leone2024monotones}.

For a mixed state described by a density matrix $\rho$, 
we furthermore define the mixed-state SRE~\cite{Leone2022}:
\begin{equation}\label{eq:Mixed state magic}
    \widetilde{M}^{(n)}(\rho)=M^{(n)}(\rho)-S^{(n)}(\rho),
\end{equation}
where $S^{(n)}(\rho)=\log(\mathrm{tr}\,\rho^n)/(1-n)$ is the $n$-th order R\'enyi entropy. Once again, for an infinite system, we are primarily interested in the associated mixed-state SRE density, $\widetilde{m}^{(n)}=\widetilde{M}^{(n)}/L$.  We note that the mixed-state SRE is generally a poor monotone for mixed-state nonstabilizerness as $\widetilde{M}^{(n)}(\rho) \neq 0$ for convex mixtures of stabilizer states \cite{tarabunga2025efficientmutualmagicmagic}. To combat this, efficient witnesses of mixed-state nonstabilizerness have been proposed to identify highly nonstabilizer mixed states \cite{haug2025efficientwitnessingtestingmagic}, which we discuss in Appendix~\ref{app:witnessing magic}.

Direct evaluation of the SRE for a chain of $L$ spins scales as $4^L$, making it intractable for large systems. This exponential scaling can be mitigated by approximating the SRE via Monte Carlo sampling in the Pauli-string basis~\cite{Tarabunga2023,lami2023quantum,Liu2025Ising}. Although this avoids the exponential cost of a direct computation, it may require a large number of samples to achieve accurate statistics. Alternatively, for MPS states, one can compute the SRE using the replica trick~\cite{Haug2023MPS}. This approach is exact even for large or infinite systems, with the computational cost determined solely by the bond dimension and the number of replicas.

The SRE of an iMPS can be calculated using a transfer matrix approach analogous to that used to evaluate standard correlation functions~\cite{Haug2023MPS}.  First, we create a $2n$-fold replica of the state $\ket{\psi}^{\otimes 2n}$ with physical dimension $d^{\prime}=d^{2n}$ and  bond dimension $\chi^{2n}$. We then define the tensors:
\begin{eqnarray}
    B_{ij}^{\sigma^{s}} = (A^\sigma_{ij})^{\otimes2n}, \quad 
    \Lambda_k^{(n)} = \frac{1}{2} \sum^{3}_{\alpha=0}(\sigma^\alpha_k \otimes \sigma^{\alpha \ast}_k)^{\otimes n},
\end{eqnarray}
where $B_{ij}^{\sigma^{s}}$ is a $2n$-fold copy of $A$, and $\Lambda_k^{(n)}$ encodes the Pauli matrices operating in replica space on a single site $k$. For ease, we will denote $\Lambda_k^{(n)}$ as $\Lambda$ for a single physical site. $\Lambda$ is then contracted over the physical bond of the tensor $B$ to create a modified $\chi^{4n}{\times}\chi^{4n}$ transfer matrix $\mathbb{E}$:
\begin{equation}\label{eq:SRETM}
\mathbb{E}_{(ik),(jl)}=\sum_{s,s^\prime} B^{\sigma^s}_{i,j}(\Lambda ^{\sigma^s,\sigma^{s^\prime}})  \bar{B}^{\sigma^{s^\prime}}_{k,l}.
\end{equation}
Denoting the dominant eigenvalue of $\mathbb{E}$ as $\mu_1$, the SRE density in the thermodynamic limit is 
\begin{equation}
 m^{(n)}=(1-n)^{-1}\log(\mu_1),   
\end{equation}
which is upper-bounded by $\log(d)$ \cite{Leone2022}.

For our analytical calculations below, we employ the above replica trick exactly. This is limited to low bond dimensions as the computation scales as $\chi^{6n}$. To access larger values of $\chi$ in the numerics, we use an equivalent method based on Pauli-basis conversion~\cite{Banuls2024}. The latter allows bond-dimension truncation throughout the computation, which improves the scaling, but introduces an approximation. We denote the bond dimension of the underlying MPS by $\chi$, and that of the truncated Pauli-MPS by $\chi_t \leq \chi^2$. Since no further truncation or variational optimization is performed, the bond dimension of the resulting $n$-th order SRE Pauli-MPS is $\chi_t^n$, and the dimension of the corresponding $\mathbb{E}$ is $\chi_t^{2n}$, which quickly becomes prohibitive to evaluate for large numbers of replicas.

The above replica trick for calculating SREs for iMPS makes it possible to study the nonstabilizerness of 1D systems directly in the thermodynamic limit. For example, Ref.~\cite{smith2024Rydbergatoms} used this approach to understand the origin of nonstabilizerness in Rydberg atom arrays and its temporal evolution following a quantum quench. Ref.~\cite{liu2025SRETIMPS} extracted the SRE for states widely used in quantum information theory and proposed a new algorithm `Bond DMRG' for calculating the nonstabilizerness for iMPS. 

\section{Nonstabilizerness of finite subsystems} \label{sec:magic_subsystems}

In Sec.~\ref{sec:overview}, we introduced the tools for calculating the SRE of an iMPS and detecting critical points in the thermodynamic limit. In particular, we discussed the role of the dominant eigenvalue of the SRE transfer matrix, $\mathbb{E}$. Building on this, we now ask: what are the roles of the subleading eigenvalues and the corresponding eigenvectors of $\mathbb{E}$? As an illustrative example, we begin by considering the SRE of finite subsystems embedded within an infinite chain and we analyze the eigendecomposition of the SRE transfer matrix. This analysis will reveal that the subleading eigenvalues give rise to a \emph{nonstabilizerness correlation length}, directly analogous to the standard correlation length extracted from the conventional MPS transfer matrix. Furthermore, the dominant eigenvector encodes the mutual SRE shared between two adjacent subsystems.

\subsection{Subleading corrections to nonstabilizerness}\label{subsec:SRE_corrections}

Analogous to the MPS transfer matrix in Eq.~(\ref{eq:TMspectral}), the SRE transfer matrix $\mathbb{E}$ can be spectrally decomposed as
\begin{equation}\label{eq:SRETMspectral}
    \mathbb{E}=
    \begin{tikzpicture}[tensornetwork, scale=1.4]
    \renewcommand{\tensorsize}{16pt}
        \node[atensor, label={center, text depth=}:\(B^{*}\)] (A3) at (3, 1) {};
        \node[atensor, label={center, text depth=}:\(B\)] (A3conj) at (3, -1) {};
        \node[ctensor, label={center, text depth=}:\(\Lambda\)] (O) at (3, 0) {};
        \draw (A3) -- (O) -- (A3conj);
        \draw (A3.west) -- +(-0.5, 0);
        \draw (A3.east) -- +(0.5, 0);
        \draw (A3conj.west) -- +(-0.5, 0);
        \draw (A3conj.east) -- +(0.5, 0);
    \end{tikzpicture}
    = \sum_{i=1}^{\chi^{4n}} \mu_{i}|R^{m}_{i})(L^{m}_{i}|,
\end{equation}
where the eigenvalues $\mu_i$ are ordered in descending magnitude and have corresponding left and right SRE eigenvectors, $(L^{m}_{i}|$ and $|R^{m}_{i})$ (we use the ``magic'' superscript $m$ to distinguish them from those of the ordinary MPS transfer matrix, $E$). The key distinction between Eq.~(\ref{eq:SRETMspectral}) and Eq.~(\ref{eq:TMspectral}) is that the leading eigenvalue $\mu_1$ only approaches unity for stabilizer states, for which the SRE vanishes. 

To compute the SRE of a finite subregion embedded in an infinite system, we restrict the sum over all Pauli strings $L$ in Eq.~(\ref{eq:SRE}) to an $N$-site subset of nontrivial Pauli strings, $\mathcal{P}_N \subseteq \mathcal{P}_L$. Within this region, each site contributes one power of the SRE transfer matrix, yielding $\mathbb{E}^N$, while outside the region all Pauli operators are trivial, so that $\Lambda \rightarrow I^{\otimes 2n}$ and the SRE transfer matrix reduces to the MPS transfer matrix, $\mathbb{E} \rightarrow E^{\otimes 2n}$. Consequently, the SRE of an $N$-site subsystem is obtained from the contraction
\begin{equation}\label{eq:SRE_N}
M^{(n)}(\rho)=(1-n)^{-1}\log(\mathbb{L}| \mathbb{E}^N |\mathbb{R}),
\end{equation}
where $(\mathbb{L}| = (L_{1}|^{\otimes 2n}$ and $|\mathbb{R}) = |R_{1})^{\otimes 2n}$ are built from the dominant eigenvectors of $E^{\otimes 2n}$, see Appendix~\ref{Appendix:SRE separated} for further details.

We can gain additional insight into  Eq.~(\ref{eq:SRE_N}) by employing the spectral decomposition of $\mathbb{E}$ introduced in Eq.~(\ref{eq:SRETMspectral}) and write $\mathbb{E}^N$ as
\begin{equation}
\mathbb{E}^{N}=
\begin{tikzpicture}[tensornetwork, scale=1.4, baseline=(current bounding box.center)]
    \renewcommand{\tensorsize}{16pt}
    
    % Nodes
    \node[atensor, label={center, text depth=}:\(B^{*}\)] (A3) at (3, 1) {};
    \node[atensor, label={center, text depth=}:\(B\)] (A3conj) at (3, -1) {};
    \node[ctensor, label={center, text depth=}:\(\Lambda\)] (O) at (3, 0) {};

    \node[atensor, label={center, text depth=}:\(B^{*}\)] (A4) at (5, 1) {};
    \node[atensor, label={center, text depth=}:\(B\)] (A4conj) at (5, -1) {};
    \node[ctensor, label={center, text depth=}:\(\Lambda\)] (O1) at (5, 0) {};
    
    % Connections
    \draw (A3) -- (O) -- (A3conj);
    \draw (A3.west) -- +(-0.5, 0);
    \draw (A3.east) -- +(0.5, 0) node[right] {\(\ldots\)};
    \draw (A3conj.west) -- +(-0.5, 0);
    \draw (A3conj.east) -- +(0.5, 0) node[right] {\(\ldots\)};
    
    \draw (A4) -- (O1) -- (A4conj);
    \draw (A4.west) -- +(-0.5, 0);
    \draw (A4.east) -- +(0.5, 0);
    \draw (A4conj.west) -- +(-0.5, 0);
    \draw (A4conj.east) -- +(0.5, 0);

    % Underbrace
    \usetikzlibrary{decorations.pathreplacing}
    \draw[decorate, decoration={brace, mirror, amplitude=5pt}]
        (2.5,-1.5) -- (5.5,-1.5)
        node[midway, yshift=-0.4cm] {$N \text{ sites}$};
\end{tikzpicture}
=
\sum_{i=1}^{\chi^{4n}} \mu_{i}^{N} |R^{m}_{i})(L^{m}_{i}|.
\end{equation}
The SRE of the $N$-site subsystem is then obtained by contracting the boundary vectors $(\mathbb{L}|$ and $|\mathbb{R})$ onto the edges of this region,
\begin{equation}\label{eq:expectation value}
    (\mathbb{L}| \mathbb{E}^N |\mathbb{R}) 
    = \sum_{i=1}^{\chi^{4n}} 
    \mu_{i}^{N}(\mathbb{L}|R^{m}_{i})(L^{m}_{i}|\mathbb{R}).
\end{equation}
To simplify notation, we define 
\begin{equation}\label{eq:c}
 c_{i}=(\mathbb{L}|P^{m}_{\mu_{i}}|\mathbb{R}),   
\end{equation}
where $P^{m}_{\mu_{i}}$ is a projector onto the eigenspace of $\mu_i$, to account for any degeneracies in the spectrum. 

For large $N$, the contribution of the leading eigenvalue $\mu_1$ dominates, so it is useful to factor it out:
\begin{equation}\label{Eq:Magic expectation value}
    (\mathbb{L}| \mathbb{E}^N |\mathbb{R})=\mu_1^N \left[c_1 + \left(\frac{\mu_2}{\mu_1}\right)^N c_2+ 
    \left(\frac{\mu_3}{\mu_1}\right)^N c_3 + \ldots \right].
\end{equation}
Thus, via Eq.~(\ref{eq:SRE_N}), the $n$-th order SRE of an $N$-site subregion is
\begin{equation} \label{eq:pure magic expansion}
\begin{split}
    M^{(n)}(\rho) %&= \frac{\text{log}(\mathbb{L}| \mathbb{E}^N |\mathbb{R})}{1-n}\\
                &=\frac{N\text{log}(\mu_1)}{1-n} +\frac{\text{log}[c_1 + f(N)]}{1-n},
\end{split}
\end{equation}
where we have defined
\begin{equation}\label{eq:f}
 f(N) \equiv \sum_{i=2}^{\chi^{4n}} \left( \frac{\mu_i}{\mu_1} \right)^N c_i\,.   
\end{equation}
In the large-$N$ limit, $\mu_{i=2,3,\ldots} / \mu_1 \rightarrow 0$, so $f(N) \ll c_1$.
We can then apply the Taylor expansion, $\log[c_1 + f(N)] \approx \log(c_1) + f(N)/c_1 +\mathcal{O}(f(N)^2/c_1^2)$, and subtract the $n$-th order Rényi entropy $S^{(n)}(\rho)$ to obtain the mixed-state SRE:
\begin{multline}\label{Eq:Magic Universal}
    \widetilde{M}^{(n)}(\rho) \approx  \frac{N\text{log}(\mu_1)}{1-n} +\frac{\text{log}(c_1)}{1-n}-S^{(n)}(\rho) + \frac{f(N)}{(1-n)c_1}\,.
\end{multline}

Ignoring higher-order terms in the Taylor expansion, Eq.~(\ref{Eq:Magic Universal}) can be organized into three distinct contributions, as advertised in Fig.~\ref{fig:equation contributions}: (i) a dominant $\mathcal{O}(N)$ extensive term determined by the leading eigenvalue $\mu_1$ of $\mathbb{E}$; (ii) an $\mathcal{O}(1)$ term (given by $c_1$ and the R\'enyi entropy), which is controlled by the overlap between the dominant eigenvectors of the SRE transfer matrix $\mathbb{E}$ and the MPS transfer matrix $E$; (iii) the correction term $f$, due to the subleading eigenvalues of $\mathbb{E}$. In the limit $N \rightarrow \infty$, $\tilde{m}^{(n)}$ converges to the contribution of the dominant eigenvalue $\mu_1$, recovering the result in Ref.~\cite{Haug2023MPS}. Finally, as we will show in the following Sec.~\ref{subsec:Long range magic proof}, the $\mathcal{O}(1)$ term can be identified as the mutual SRE between two equal-sized adjacent subsystems in the thermodynamic limit.

Since $f(N)$ is determined by the subleading eigenvalues of $\mathbb{E}$ in Eq.~(\ref{eq:f}), it vanishes as $N \rightarrow \infty$. In this limit, the dominant contribution to Eq.~(\ref{eq:f}) arises from the second-largest eigenvalue, giving 
\begin{equation}\label{eq:subleading SRE term}
f(N) \underset{N\to\infty}{\longrightarrow} c_2 e^{-N / \xi^{(n)}_{\mathrm{SRE}}},  \;\;\;\; \xi^{(n)}_{\mathrm{SRE}} = \frac{-1}{\log(|\mu_2 / \mu_1|)},
\end{equation}
where $\xi^{(n)}_{\mathrm{SRE}}$ is the \emph{SRE correlation length}. This correlation length characterizes the longest-range correlations present in the SRE of an MPS and is therefore a direct analog to the standard MPS correlation length, Eq.~(\ref{Eq:correlation length}). In general, $\xi^{(n)}_{\mathrm{SRE}}$ and the conventional MPS correlation length $\xi$ can differ quantitatively from one another, reflecting the distinct structure of nonstabilizer correlations. However, as we will demonstrate in Sec.~\ref{sec:MPS_skeleton} and Sec.~\ref{sec:fullclusterIsing}, $\xi^{(n)}_{\mathrm{SRE}}$ exhibits diverging behavior similar to the ordinary correlation length at quantum critical points, suggesting that it may serve as a universal indicator of criticality in quantum many-body systems. Finally, $\xi^{(n)}_{\mathrm{SRE}}$ also governs the decay of the SRE when a state is perturbed on two spatially separated sites, as we will demonstrate in Sec.~\ref{sec:quantum_circuits}.

\subsection{Mutual SRE}\label{subsec:Long range magic proof}

Having clarified the physical interpretation of the extensive and exponentially decaying components of Eq.~(\ref{Eq:Magic Universal}), we now turn to its $\mathcal{O}(1)$ term. We will show that this contribution is directly related to the mutual SRE between two adjacent subsystems $A$ and $B$~\cite{Tarabunga2023,hoshino2025stabilizerrenyientropyconformal},
\begin{equation}\label{eq:mutual magic}
L^{(n)}(A\,{:}\,B)=\widetilde{M}^{(n)}(\rho_A)+\widetilde{M}^{(n)}(\rho_B)-\widetilde{M}^{(n)}(\rho_{AB}).
\end{equation}
The quantity $L^{(n)}(A\,{:}\,B)$ measures the degree of nonstabilizerness in the correlations between the two subsystems, analogous to the mutual information~\cite{Nielsen_Chuang_2010,Melko2010montecarloMI,Wilms_2011Mutualinfo,Calabrese2004,Kudler2023,Wolf2008MI}. Moreover, it quantifies the nonstabilizerness that cannot be removed by local unitary rotations~\cite{White_2021}. Owing to the additive nature of the SRE, the mutual SRE vanishes when $\rho = \rho_A \otimes \rho_B$ is a product state, although it does not generally satisfy the additivity property of the mutual information. Related quantities have also been introduced for other measures of nonstabilizerness, such as the robustness of magic~\cite{Sarkar_2020,Bao2022} and the mana~\cite{White_2021,Fliss2021}.

The mutual SRE is defined using the mixed state SRE in Eq.~(\ref{eq:Mixed state magic}), therefore it can be decomposed according to $L^{(n)}(A\,{:}\,B)=W^{(n)}(A\,{:}\,B)-I^{(n)}(A\,{:}\,B)$, where
\begin{equation}
    W^{(n)}(A\,{:}\,B)=M^{(n)}(\rho_A)+M^{(n)}(\rho_B)-M^{(n)}(\rho_{AB})
\end{equation}
 is the pure state mutual SRE, and
\begin{equation}
     I^{(n)}(A\,{:}\,B)=S^{(n)}(\rho_A) +S^{(n)}(\rho_B)-S^{(n)}(\rho_{AB})
\end{equation}
is the mutual information between the two subsystems.

We now consider the SRE of two adjacent subsystems, both of size $\ell$. Since the MPS is translation-invariant, we know that $M^{(n)}(\rho_A)=M^{(n)}(\rho_B)$, therefore the mutual SRE will be $W^{(n)}(A\,{:}\,B)=2M^{(n)}(\rho_A)-M^{(n)}(\rho_{AB})$ and $I^{(n)}(A\,{:}\,B)=2S^{(n)}(\rho_A)-S^{(n)}(\rho_{AB})$.  Using Eq.~(\ref{eq:pure magic expansion}), the extensive terms cancel, so the mutual SRE becomes:
\begin{equation}
     W^{(n)}(A\,{:}\,B)= \frac{2\log[c_1 + f(\ell)]}{1-n}
        - \frac{\log[c_1 + f(2\ell)]}{1-n}.     
\end{equation}
As $\ell \rightarrow \infty$ and the two subsystems are grown toward the thermodynamic limit, we have $f(\ell) \rightarrow 0$ and $f(2\ell) \rightarrow 0$, since these terms encode the exponentially-decaying correlations identified in Eq.~(\ref{eq:subleading SRE term}). Moreover, because $|\psi(A)\rangle$ is an MPS and therefore obeys an entanglement area law, it follows that the R\'enyi entropy of $\rho_{AB}$ saturates to a boundary contribution identical to that of a single semi-infinite region. Hence, $S^{(n)}(\rho_{AB}) \to S^{(n)}(\rho_{A/B})$ and $I^{(n)}(A\,{:}\,B) \to S^{(n)}(\rho_{A})$ as $\ell\to\infty$. In this limit, the mutual SRE simplifies to
\begin{equation}\label{Eq:mutual SRE definition}
L^{(n)}_{\infty}
= \frac{\log(c_1)}{1-n} - S^{(n)}(\rho_{A}) \equiv W^{(n)}_{\infty}- S^{(n)}(\rho_{A}).
\end{equation}
This expression corresponds precisely to the $\mathcal{O}(1)$ term identified in Eq.~(\ref{Eq:Magic Universal}). Note that $L^{(n)}_{\infty}$ is not required to be strictly positive, as the mutual SRE -- unlike the mutual information $I^{(n)}(A\,{:}\,B)$ -- does not obey the subadditivity condition. 

\subsection{Relation to boundary CFT results}

A natural question is how our iMPS expressions derived above relate to the universal content of the SRE predicted by the boundary conformal field theory (BCFT)
analysis of Ref.~\cite{hoshino2025stabilizerrenyientropyconformal} and the exact microscopic result available for free fermions~\cite{rajabpour2025stabilizershannonrenyiequivalenceexact}.  The BCFT formalism assumes that the system is at a
conformally-invariant fixed point and derives the size dependence of the SRE directly from the replicated boundary theory.  In this
approach, the SRE of an $L$-site critical pure state on a ring with periodic boundary conditions takes the form
\begin{equation}\label{eq:BCFT1}
 M^{(n)}(L)=m^{(n)} L - c_n + o(1),   
\end{equation}
where the extensive coefficient
$m^{(n)}$ is nonuniversal, while the $O(1)$ term $c_n$ is universal
and determined by the $g$-factor of the $2n$-replicated BCFT. For open boundary conditions, there is an additional $\log L$ contribution to Eq.~(\ref{eq:BCFT1})~\cite{rajabpour2025stabilizershannonrenyiequivalenceexact}. Likewise, the mutual SRE of a subsystem of $\ell$ spins exhibits the universal logarithmic scaling
\begin{equation}\label{eq:BCFT2}
       W^{(n)}(\ell)=\frac{4\Delta_{2n}}{n-1}\,\log \ell_c\,,
\end{equation}
with $\Delta_{2n}$ the scaling dimension of the relevant boundary-condition–changing operator~\cite{hoshino2025stabilizerrenyientropyconformal} and $\ell_c=(L/\pi)\sin(\pi \ell/L)$ the associated chord length. In special cases, such as the transverse-field Ising chain, $W^{(2)}(\ell)$ can obey scaling identical to the mutual information $I^{(2)}(\ell)$, leading to no logarithmic dependence of $L^{(2)}(\ell)$ \cite{hoshino2025stabilizerrenyientropyconformal}.

Our iMPS framework arrives at expressions with the same structure via a microscopic route. The spectral decomposition of the SRE transfer matrix leads to
the three-term structure in Eq.~\eqref{Eq:Magic Universal}: an
extensive contribution proportional to $\log \mu_1$, an $O(1)$ term
controlled by the overlap $c_1$ of the dominant eigenvectors, and
subleading corrections $f(N)$ that decay with the SRE correlation length
$\xi^{(n)}_{\mathrm{SRE}}$.  This decomposition is the lattice analog  of the BCFT result, Eq.~(\ref{eq:BCFT1}).  Because the replica
transfer matrix is built from $2n$ copies of the MPS tensor, both $c_1$
and the prefactor $(1-n)^{-1}$ carry explicit $n$-dependence, mirroring
the $n$-dependence of the BCFT boundary constant $c_n$. The correspondence becomes especially transparent for the mutual SRE.
Our thermodynamic-limit
expression in Eq.~(\ref{Eq:mutual SRE definition}) isolates the $O(1)$ boundary contribution, directly
analogous to the BCFT result in Eq.~(\ref{eq:BCFT2}), where $W_n(\ell)$ is governed by the
two-point function of boundary-condition–changing operators and acquires
a universal logarithmic dependence on subsystem size with coefficient
fixed by $\Delta_{2n}$.  In the iMPS setting, the same behavior emerges on length scales $\xi^{(n)}_{\mathrm{SRE}} \ll \ell$, where the
finite-bond-dimension state faithfully approximates the conformal fixed
point.

Finally, Ref.~\cite{hoshino2025stabilizerrenyientropyconformal} did not discuss the subleading corrections to Eq.~(\ref{eq:BCFT1}) which, according to our analysis, are crucial for defining the SRE correlation length, $\xi_\mathrm{SRE}^{(n)}$. As we will further demonstrate in the following sections, the
conventional correlation length $\xi$ and the SRE correlation length
$\xi^{(n)}_{\mathrm{SRE}}$ may diverge with different exponents at a critical point. This is unsurprising since, from the
field-theoretical perspective, $\xi$ and $\xi^{(n)}_{\mathrm{SRE}}$
correspond to inverse gaps of different operators in distinct replicated
theories and therefore are not \emph{a priori} expected to share the same critical
exponent.

\section{Nonstabilizer correlations due to local perturbations}\label{sec:quantum_circuits}

In Sec.~\ref{sec:magic_subsystems}, we showed that the approach of the SRE of a finite $N$-site subsystem to its thermodynamic limit value is determined by exponentially-decaying subleading terms governed by the SRE correlation length $\xi^{(n)}_{\mathrm{SRE}}$, which closely resembles the standard MPS correlation length $\xi$. The latter governs the decay of connected correlators, Eq.~(\ref{eq:arealaw}), as the distance $r$ between observables is varied. In this section, we argue that $\xi^{(n)}_{\mathrm{SRE}}$ quantifies how the SRE changes when a wave function is perturbed on two spatially separated sites.

We begin by considering the effect of locally perturbing an MPS wave function on a single site. We apply an arbitrary unitary matrix $\hat{U}_i$ to the site $i$ of a translationally-invariant iMPS, $|\psi(A)\rangle \rightarrow \hat{U}_i |\psi(A)\rangle$, which corresponds to a modification of the MPS tensor $A^{\sigma_i}$ at that site:
\begin{equation}
(A_{U})^{\sigma^\prime_i}_{\alpha,\beta} = \sum_{\sigma_i} U^{\sigma_i^{\prime},\sigma_i} A^{\sigma_i}_{\alpha,\beta}.
\end{equation}
The modified tensor $A_U$ can be projected into the replica space by constructing a $2n$-fold tensor product,
$(B_U)_{\alpha,\beta}^{\sigma_i} = ((A_{U})_{\alpha,\beta}^{\sigma_i})^{\otimes 2n}$,
with bond dimension $\chi^{2n}$, physical dimension $d^{2n}$, and an associated transfer matrix $\mathbb{E}_U$ by analogy with $\mathbb{E}$ in Eq.~(\ref{eq:SRETM}).

The effect of a local perturbation on the SRE can be captured by replacing the $\mathbb{E}$ transfer matrix on site $i$ with the modified SRE transfer matrix $\mathbb{E}_U$, while leaving $\mathbb{E}$ unchanged elsewhere:
\begin{equation} \label{eq:oneunitary}
\begin{tikzpicture}[tensornetwork, scale=1.4, baseline=(current bounding box.center)]
    \renewcommand{\tensorsize}{16pt}
    
    % Nodes

    \node[atensor,draw=none, label={center, text depth=}:\( \ldots \)] (A1) at (1.3, 0) {};

    \node[atensor,draw=none, label={center, text depth=}:\( \ldots \)] (A6) at (6.7, 0) {};

    \node[atensor, label={center, text depth=}:\(B^{*}\)] (A2) at (2, 1) {};
    \node[atensor, label={center, text depth=}:\(B\)] (A2conj) at (2, -1) {};
    \node[ctensor, label={center, text depth=}:\(\Lambda\)] (O2) at (2, 0) {};
    
    \node[atensor, label={center, text depth=}:\(B^{*}\)] (A3) at (3, 1) {};
    \node[atensor, label={center, text depth=}:\(B\)] (A3conj) at (3, -1) {};
    \node[ctensor, label={center, text depth=}:\(\Lambda\)] (O3) at (3, 0) {};

    \node[atensor, label={center, text depth=}:\(B_{U}^{*}\)] (A4) at (4, 1) {};
    \node[atensor, label={center, text depth=}:\(B_{U}\)] (A4conj) at (4, -1) {};
    \node[ctensor, label={center, text depth=}:\(\Lambda\)] (O4) at (4, 0) {};

    \node[atensor, label={center, text depth=}:\(B^{*}\)] (A5) at (5, 1) {};
    \node[atensor, label={center, text depth=}:\(B\)] (A5conj) at (5, -1) {};
    \node[ctensor, label={center, text depth=}:\(\Lambda\)] (O5) at (5, 0) {};
    
    \node[atensor, label={center, text depth=}:\(B^{*}\)] (A6) at (6, 1) {};
    \node[atensor, label={center, text depth=}:\(B\)] (A6conj) at (6, -1) {};
    \node[ctensor, label={center, text depth=}:\(\Lambda\)] (O6) at (6, 0) {};
    
    % Connections
    \draw (A2) -- (O2) -- (A2conj);
    \draw (A2.west) -- +(-0.5, 0);
    \draw (A2.east) -- +(0.25, 0) ;
    \draw (A2conj.west) -- +(-0.5, 0);
    \draw (A2conj.east) -- +(0.25, 0) ;
    
    \draw (A3) -- (O3) -- (A3conj);
    \draw (A3.west) -- +(-0.25, 0);
    \draw (A3.east) -- +(0.25, 0) ;
    \draw (A3conj.west) -- +(-0.25, 0);
    \draw (A3conj.east) -- +(0.25, 0) ;

    \draw (A4) -- (O4) -- (A4conj);
    \draw (A4.west) -- +(-0.25, 0);
    \draw (A4.east) -- +(0.25, 0) ;
    \draw (A4conj.west) -- +(-0.25, 0);
    \draw (A4conj.east) -- +(0.25, 0) ;
    
    \draw (A5) -- (O5) -- (A5conj);
    \draw (A5.west) -- +(-0.25, 0);
    \draw (A5.east) -- +(0.25, 0);
    \draw (A5conj.west) -- +(-0.25, 0);
    \draw (A5conj.east) -- +(0.25, 0);

    \draw (A6) -- (O6) -- (A6conj);
    \draw (A6.west) -- +(-0.25, 0);
    \draw (A6.east) -- +(0.5, 0) ;
    \draw (A6conj.west) -- +(-0.25, 0);
    \draw (A6conj.east) -- +(0.5, 0) ;

\end{tikzpicture}
\end{equation}
This contraction can be evaluated by replacing the infinite environment of $\mathbb{E}$ with its dominant left and right eigenvectors, $(L^{m}_1|$ and $|R^{m}_1)$, and including the leading eigenvalue $\mu_1$ as a prefactor:
\begin{equation}
    \label{eq:oneunitary_contracted}    \mu_1^\infty(L^{m}_1|\mathbb{E}_U|R^{m}_1)= \mu_1^\infty
    \begin{tikzpicture}[tensornetwork, scale=1.6]
    \renewcommand{\tensorsize}{16pt}
        % --- Nodes ---
        \node[ctensor, label={center, text depth=}:\(L^{m}_{1}\)] (L) at (0, 0) {};
        \node[atensor, label={center, text depth=}:\(B_U^{*}\)] (Bconj) at (1, 1) {};
        \node[ctensor, label={center, text depth=}:\(\Lambda\)] (Lambda) at (1, 0) {};
        \node[atensor, label={center, text depth=}:\(B_U\)] (B) at (1, -1) {};
        \node[ctensor, label={center, text depth=}:\(R^{m}_1\)] (R) at (2, 0) {};

        % --- Connections ---
        \draw (Bconj) -- (Lambda) -- (B);
        \draw[rounded corners] (Bconj.east) -- (Bconj.east -| R.north) -- (R)
              -- (R.south |- B.east) -- (B.east);
        \draw[rounded corners] (Bconj.west) -- (Bconj.west -| L.north) -- (L)
              -- (L.south |- B.west) -- (B.west);
    \end{tikzpicture}.
\end{equation}
Since we are considering iMPS, the total SRE contains an extensive (and hence divergent) contribution from the dominant eigenvalue $\mu_1$. However, we are interested only in the change in SRE due to the  perturbation:
\begin{equation}
\delta M^{(n)}_U = M^{(n)}(\hat{U}_i|\psi\rangle) - M^{(n)}(|\psi\rangle),
\end{equation}
for which this extensive contribution cancels.
To isolate this finite difference, we rescale $\mathbb{E}$ and $\mathbb{E}_U$ by the dominant eigenvalue $\mu_1$, defining $\tilde{\mathbb{E}} \equiv  \mathbb{E}/\mu_1$ and $\tilde{\mathbb{E}}_U \equiv  \mathbb{E}_U/\mu_1$, such that
$(L^m_1|\tilde{\mathbb{E}} = (L^m_1|$ and $\tilde{\mathbb{E}}|R^{m}_1) = |R^{m}_1)$.
The resulting change in the SRE is then
\begin{equation}
\label{eq:oneunitary_SRE}
\delta M^{(n)}_U = \frac{\log{} (L^{m}_1|\tilde{\mathbb{E}}_U|R^{m}_1)}{1-n}
= \frac{\log{} (L^{m}_1|\mathbb{E}_U|R^{m}_1) - \log \mu_1}{1-n}.
\end{equation}

Having found the change in the SRE under a single local perturbation, we now consider the effect of two spatially separated local perturbations on the SRE as we apply local unitary operators $\hat{U}_{i}$ and $\hat{U}_{i+r}$ to the state, $|\psi(A)\rangle \rightarrow \hat{U}_i \hat{U}_{i+r}|\psi(A)\rangle$. Similar to the above, we wish to compute the relative change in the SRE:
\begin{equation}
\delta M^{(n)}_{U,U} = M^{(n)}(\hat{U}_i\hat{U}_{i+r}|\psi\rangle) - M^{(n)}(|\psi\rangle).
\end{equation}
Analogous to Eq.~\eqref{eq:oneunitary} and Eq.~\eqref{eq:oneunitary_contracted}, this quantity can be evaluated by the following contraction:
\begin{equation}
(L^{m}_1|\tilde{\mathbb{E}}_{U}
\tilde{\mathbb{E}}^{r-1} \tilde{\mathbb{E}}_{U}|R^{m}_1).
\end{equation}
To evaluate this, we separate the dominant eigenvalue component of the modified transfer matrix $\tilde{\mathbb{E}}$ from the subleading, exponentially decaying terms, writing
$\tilde{\mathbb{E}} = |R^{m}_1)(L^{m}_1| + \left(\tilde{\mathbb{E}} - |R^{m}_1)(L^{m}_1|\right)$.
Substituting this decomposition gives
\begin{equation} \begin{split} &(L^{m}_1|\tilde{\mathbb{E}}_{U} \tilde{\mathbb{E}}^{r-1} \tilde{\mathbb{E}}_{U}|R^{m}_1)=(L^{m}_1|\tilde{\mathbb{E}}_{U} |R^{m}_1)(L^{m}_1| \tilde{\mathbb{E}}_{U}|R^{m}_1)\\&+\sum_{i}^{\chi^{4n}}\left(\frac{\mu_i}{\mu_1}\right)^{r-1}(L^{m}_1|\mathbb{\tilde{E}}_{U}|R^{m}_{i})(L^{m}_{i}| \mathbb{\tilde{E}}_{U}|R^{m}_1), \end{split} \end{equation}
which contains a constant term, equal to the square of Eq.~\eqref{eq:oneunitary_contracted}, and an exponentially-decaying correction. This structure is directly analogous to the behavior of two-point correlators in iMPS, although here the logarithmic definition of the SRE modifies the dependence slightly. 

Introducing the notation
\begin{equation}
 d_i = (L^{m}_1|\mathbb{\tilde{E}}_{U}|R^{m}_{i})(L^{m}_{i}|\mathbb{\tilde{E}}_{U}|R^{m}_1),   
\end{equation}
and assuming that $r$ is large enough that $1 \gg \mu_2/\mu_1 \gg \mu_3/\mu_1$,
we obtain
\begin{equation} \begin{split} \delta M^{(n)}_{U,U}&\approx\frac{1}{1-n}\text{log}\left(d_1+d_2e^{-r/\xi^{(n)}_{\mathrm{SRE}}}\right) \\ &\approx\frac{\log{d_1}}{1-n}+\frac{(d_2/d_1)e^{-r/\xi^{(n)}_{\mathrm{SRE}}}}{1-n}\\ &=2\delta M^{(n)}_U+\frac{d_2}{d_1(1-n)}e^{-r/\xi^{(n)}_{\mathrm{SRE}}}\,. \end{split} \end{equation}
Hence, the change in the SRE due to two spatially-separated unitary perturbations consists of two distinct contributions: (i) a disconnected term equal to twice the single-unitary result of Eq.~\eqref{eq:oneunitary_SRE}, and
(ii) an exponentially decaying term controlled by the SRE correlation length $\xi^{(n)}_{\mathrm{SRE}}$. Now it becomes transparent why $\xi^{(n)}_{\mathrm{SRE}}$ can generally differ from the conventional MPS correlation length: applying a local unitary affects not only the expectation values of nearby single-site Pauli operators but also those of multi-site Pauli strings. The emergent nonlocal correlations among these operators are precisely captured by the SRE correlation length $\xi^{(n)}_{\mathrm{SRE}}$.

\begin{figure*}
    \centering
    \includegraphics[width=0.99\linewidth]{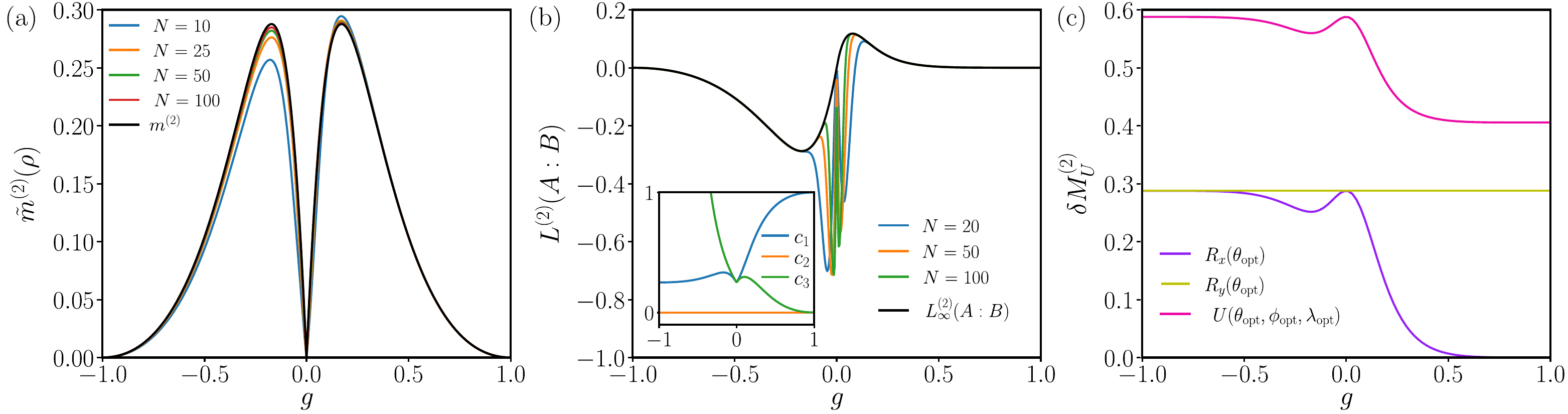}
    \caption{Nonstabilizer properties of the MPS skeleton in Eq.~(\ref{Eq:Skelton MPS}). (a) The mixed-state SRE density $\tilde{m}^{(2)}$ over a subsystem of increasing size $N$, illustrating convergence to the thermodynamic limit value in Eq.~(\ref{eq:skeleton peak}) (black line labeled $m^{(2)}$). (b) The mutual SRE, $L^{(2)}(A\,{:}\,B)$ in Eq.~\eqref{eq:mutual magic}, between two connected subsystems $A$ and $B$ of size $N$. Black line shows the thermodynamic limit value, $L^{(2)}_{\infty}$ [Eq.~(\ref{Eq:mutual SRE definition}) with Eqs.~(\ref{eq:S2skeleton})-(\ref{eq:skeletonmutualanalytic})]. Inset: Correction coefficients $c_i$ in the decomposition of the nonstabilizerness. (c): The maximum change in the SRE, $\delta M^{(2)}_U$, caused by an application of a single-qubit unitary. The optimal angles $\theta,\phi,\lambda$ were found by numerical optimization of Eq.~\eqref{eq:oneunitary_SRE}. The finite-size results in panels (a)-(b) were obtained by numerical evaluation of Eq.~\eqref{eq:Mixed state magic} and Eq.~\eqref{eq:mutual magic}.} 
    \label{fig:Skeleton SRE Decay}
\end{figure*}

\section{Nonstabilizerness of the MPS skeleton} \label{sec:MPS_skeleton}

As an analytically tractable example, we apply the formalism developed in the previous sections to a simple, low bond-dimension MPS ansatz that contains a quantum phase transition between a product phase and a symmetry-protected topological phase (SPT). We choose a single-parameter path through the spin-$1/2$ cluster-Ising model whose ground state can be exactly represented by an MPS~\cite{Wolf2006Skeleton,Smith2022Ctopologicalphase}:
\begin{eqnarray}
    \label{eq:MPS skeleton Hamiltonian}
    \notag H_{\text{skeleton}} &=& \sum_{i} (g-1)^2\sigma^{z}_{i}\sigma^{x}_{i+1}\sigma^{z}_{i+2} +2(g^2 -1)\sigma^{z}_{i}\sigma^{z}_{i+1} \\
     &-& (1+g)^{2}\sigma^{x}_{i},    
\end{eqnarray}
where $g =-1$ reduces to the pure cluster Hamiltonian and $g = 1$ corresponds to the free paramagnet. As we will see below,  $g = 0$ is special -- it represents a ``multicritical'' point where a trivial paramagnet, symmetry-broken Ising phase, and an SPT (cluster) phase meet. 

The path traced out by $g$ in Eq.~(\ref{eq:MPS skeleton Hamiltonian}) is contained in a larger family of solvable models dubbed the ``MPS skeleton'', which describe a class of SPT phases whose ground states are exactly represented by a finite bond dimension MPS~\cite{Jones2021Skeletons}. The $\chi=2$ skeleton that captures the ground state of the Hamiltonian in Eq.~(\ref{eq:MPS skeleton Hamiltonian}) is given by matrices:
\begin{equation}\label{Eq:Skelton MPS}
    A^{\uparrow}=\begin{pmatrix}
        0&0\\
        1&1
    \end{pmatrix},\;\;\;
    A^{\downarrow}=\begin{pmatrix}
        1&g\\
        0&0
    \end{pmatrix}.
\end{equation}
It is curious that this MPS skeleton describes exactly the ground state
even at the critical point $g = 0$, despite the fact that the entanglement entropy there is strictly bounded by $\log{2}$. This is because $g=0$ is a multicritical point at which the entanglement entropy does not diverge with subsystem size, unlike in second-order  transitions. The latter are also present in the full cluster-Ising model for other choices of couplings and will be the subject of Sec.~\ref{sec:fullclusterIsing}. In Appendix~\ref{Appendix:Higher order skeleton}, we study an example of a higher order MPS skeleton with $\chi=4$, showing that our approach directly generalizes to that case. 

\subsection{Nonstabilizer properties}

Since the MPS skeleton in Eq.~(\ref{Eq:Skelton MPS}) has a small bond dimension $\chi=2$, it is possible to analytically understand the behavior of its nonstabilizerness for $n=2$ replicas using \textit{Mathematica}. The (unnormalized) eigenvalues of the standard transfer matrix, Eq.~(\ref{eq:TMspectral}), are given by:
$\lambda_1=1+g$, $\lambda_2=1-g$, $\lambda_{3,4}=0$. Moreover, the eigenvalues of the $n=2$ SRE transfer matrix $\mathbb{E}$ in Eq.~(\ref{eq:SRETM}) are:
\begin{equation}\label{eq:Skeleton eigenvalues}
    \begin{split}
        &\mu_{1}=\frac{1+14g^2 + g^4}{(1+|g|)^4} \text{ , } \mu_{2,...,9}= \frac{1-g^4}{(1+|g|)^4}\\
        &\mu_{10,...,16}=\frac{(g^2 - 1)^2} {(1+|g|)^4} \text{ , }\mu_{17,...,256}=0.
    \end{split}
\end{equation}
In the thermodynamic limit, the SRE density is determined by the dominant eigenvalue $\mu_1$ and takes the closed form expression: 
\begin{equation}\label{eq:skeleton peak}
    m^{(2)}=-\log\frac{1+14g^2 + g^4}{(1+|g|)^4}.
\end{equation}
We note that this formula for $g>0$ is equivalent to the SRE of the ground state of the 1D stochastic mean-field Ising model using the substitution g=$e^{-2\beta}$~\cite{Tarabunga2024RK}.

In Fig.~\ref{fig:Skeleton SRE Decay}(a) we evaluate the mixed-state SRE density $\tilde{m}^{(2)}(\rho)$ for a subsystem of size $N$ of the MPS skeleton and study how it approaches the SRE value in the thermodynamic limit, Eq.~\eqref{eq:skeleton peak}. To calculate the $S^{(2)}(\rho)$ contribution for finite subsystems, we utilize swap tricks~\cite{Zou2016MPSswaptick,Hastings2010Renyiswaptrick} between two replicas of the subsystem tensors. In the SPT phase ($g<0$), we find that $\tilde{m}^{(2)}(\rho)$ converges more slowly than in the paramagnet  phase, and Eq.~\eqref{eq:skeleton peak} serves as an upper bound. This is because the SPT ground state is entangled, allowing the $S^{(2)}(\rho)$ term to dominate and lowering $\tilde{m}^{(2)}(\rho)$ below the thermodynamic limit value in Eq.~\eqref{eq:skeleton peak}. 

Furthermore, in Fig.~\ref{fig:Skeleton SRE Decay}(a) we see that for $g=-1,0,1$, the SRE vanishes, implying the ground state at these points is a stabilizer state. Indeed, for $g=-1$, the ground state is the cluster state, which we know can be represented as a graph state and hence a stabilizer state. For $g=+1$, the paramagnet ground state is a product state, hence also a stabilizer state. Finally, at the multicritical point $g=0$ the ground state is a Greenberger-Horne-Zeilinger (GHZ) state \cite{Wolf2006Skeleton}, which is also a stabilizer state since it can be prepared using only Clifford gates. 

A surprising feature of Fig.~\ref{fig:Skeleton SRE Decay}(a) is the peak of nonstabilizerness. From Eq.~\eqref{eq:skeleton peak}, it is easy to show that the peak occurs at
\begin{equation}\label{eq:gstar}
 g_*=\pm(3-2\sqrt{2}), \quad m_{*}^{(2)}\approx0.28,
\end{equation}
which is comparable to the SRE of the Ising model at its critical point~\cite{Tarabunga2023,Liu2025Ising}. 
To explain this SRE peak, we follow the approach in Ref.~\cite{smith2024Rydbergatoms} and consider the unitary which prepares the MPS skeleton state by acting on the reference state $\ket{0}$. Ref.~\cite{Smith2022Ctopologicalphase} showed that this MPS can be embedded into a 2-qubit unitary, with one site acting on $\ket{0}$, and decomposed into the quantum circuit:
\begin{equation}
            \begin{quantikz}
                         &  \octrl{1}  &  \ctrl{1}  &  \gate{X}  &\qw\\
       \ket{0} &  \gate{U(\theta_w)}   &  \gate{U(\theta_v)}  &  \hphantomgate{} &\qw 
    \end{quantikz}
\end{equation}
 where the unitaries $U(\theta)$,
 \begin{equation}
    U(\theta)=\begin{pmatrix}
        \text{sin }\theta & \text{cos }\theta\\
        \text{cos }\theta & -\text{sin }\theta
    \end{pmatrix},
\end{equation}
are parametrized by angles 
\begin{equation}
\theta_{v}=\arcsin\left(\frac{\sqrt{|g|}}{\sqrt{1+|g|}}\right), \quad \theta_{w}=\arccos\left(\frac{\mathrm{sgn}(g)\sqrt{|g|}}{\sqrt{1+|g|}}\right).
\end{equation}
The quantum circuit above only has two non-Clifford gates, hence the maximum nonstabilizerness is achieved iff  $U(\theta_v)$ and $U(\theta_w)$ are maximally non-Clifford. This occurs when $\theta_{v}=\pm\pi/8$ and $\theta_{w}=\pm3\pi/8$, i.e., precisely at  $g_{*}$ given by Eq.~(\ref{eq:gstar}). 

To evaluate the mutual SRE, we require the R\'enyi entropy from outside the two embedded subsystems, which we can extract from the canonicalized dominant right eigenvector~\cite{Smith2022Ctopologicalphase}:
\begin{equation}\label{eq:S2skeleton}
    S^{(2)}(\rho)=\Biggl\{ \begin{array}{ccc}
        2\text{log}\frac{2(1+g)^2}{1+6g+g^2} & \text{if}&g\geq0, \\
        \text{log(4)}& \text{if} &g<0.
    \end{array}
\end{equation}
The coefficient given by the overlap of the boundary vectors with $(L^{m}_1|$ and $|R^{m}_1)$, Eq.~(\ref{eq:c}), used for the calculation of $L^{(2)}_{\infty}$ is:
\begin{equation}\label{eq:skeletonmutualanalytic}
        c_1=\Biggl\{ \begin{array}{ccc}
        \frac{(1+g(4+g(22+g(4+g))))^4}{4(1+g)^4(1+14g^2+g^4)} & \text{if}&g\geq0, \\
        \frac{(-1+g)^4}{1+14g^2+g^4} & \text{if} &g<0.
    \end{array}
\end{equation}

In Fig.~\ref{fig:Skeleton SRE Decay}(b) we plot the mutual SRE, $L^{(2)}(A\,{:}\,B)$ defined in Eq.~\eqref{eq:mutual magic}, of two semi-infinite adjacent subsystems. We compare $L^{(2)}(A\,{:}\,B)$ of various system sizes to the analytic prediction given in Eq.~\eqref{Eq:mutual SRE definition}. Far away from the multicritical point ($g=0$) we obtain perfect agreement with $L^{(2)}_{\infty}$. As we approach the multicritical point, the mutual SRE deviates heavily from the analytical prediction, implying a strong dependence on system size. As we will see below, this is due to the diverging correlation length $\xi^{(2)}_{\mathrm{SRE}}\to\infty$ at the multicritical point. In the product phase, $L^{(2)}_\infty$ is positive whereas in the SPT phase it is negative. The negativity is due to the SPT phase being entanglement-dominated, with entanglement entropy saturating the upper bound $S^{(2)}(\rho)=\log(2)$, while the product phase has lower entanglement (vanishing as $g\rightarrow1$) and allowing the magic to dominate. 

In the inset of Fig.~\ref{fig:Skeleton SRE Decay}(b) we illustrate the behavior of $c_i$, Eq.~(\ref{eq:c}), for the MPS skeleton. The trivial phase exhibits strong dependence on $c_1$, up to the multicritical point $g=0$ where $c_1$ and $c_3$ become degenerate. Along the entire trajectory, we find that there is no dependence on $c_2$ for this choice of boundary vectors. In the SPT phase, $c_3$ diverges while $c_1 \rightarrow0.25$ as $g\rightarrow-1$. These divergences are acceptable because the coefficients $c_i$ are weighted with the corresponding eigenvalue, which provides an effective regularization.

Finally, it is possible to calculate analytically the change in SRE, $\delta M_{U}^{(2)}$, due to the application of a rotation gate along a single axis. Using the standard spin rotation matrix $R_x(\theta) = \exp(-i\theta\sigma_x/2)$ (and similarly for $y$ and $z$-rotations), the change in the SRE of the MPS skeleton after applying a rotation by $\theta$ along the $y$ or $z$-axis is given by:
\begin{equation}
\label{eq:RyRz rotation}
    \delta M^{(2)}_{R_y}(\theta) = \delta M^{(2)}_{R_z}(\theta) = \log\frac{8}{7+\text{cos}(4\theta)}\,,
\end{equation}
while the rotation along the $x$-axis yields the following lengthy expression:
\begin{eqnarray} \label{eq:Rx rotation}
  \nonumber      \delta M^{(2)}_{R_x}(\theta,g) &=& \text{log}(8) +2\text{log}(1+14g^2 + g^4) \\
   \nonumber     &-& \log\Big[7+212g^2 +64g^3 + 1482g^4 +64g^5 \\
   \nonumber     &+& 212g^6 + 7g^8 + \Big(1+12g^2-64g^3 \\
        &+& 102g^4 - 64g^5 +12g^6 +g^8\Big)\text{cos}(4\theta)\Big].
\end{eqnarray}
Now it is important to ask: what is the maximum amount of nonstabilizerness we can inject into the MPS skeleton solely by single-qubit unitary operations? This question can be answered by maximizing the change in SRE for different single-qubit rotation matrices. This basis-invariant perspective is closely related to recent efforts to isolate the genuinely non-local component of nonstabilizerness by removing local unitary contributions~\cite{cao2025gravitationalbackreactionmagical,Qian2025,Cepollaro2025,ahmad2025experimentaldemonstrationnonlocalmagic}.

In Fig.~\ref{fig:Skeleton SRE Decay}(c) we plot the maximum change in the SRE, $\delta M^{(2)}_U$, obtained by applying an arbitrary rotation $U(\theta,\phi,\lambda)=R_{z}(\phi)R_{y}(\theta)R_z(\lambda)$. We calculate the maximum $\delta M^{(2)}_U$ by numerically optimizing Eq.~\eqref{eq:oneunitary_SRE} to find the optimal angles $\{\theta_{\text{opt}},\phi_{\text{opt}},\lambda_{\text{opt}}\}$. For an $R_{y,z}(\theta)$ rotation we find, consistent with Eq.~\eqref{eq:RyRz rotation}, that $\delta M^{(2)}_{U}$ is independent of $g$ and therefore maximized when $\theta_{\text{opt}}=3\pi/4$ to give $\delta M^{(2)}_{R_z}=\text{log}(4/3)$ for all $g$. For an $R_{x}(\theta)$ rotation, from Eq.~\eqref{eq:Rx rotation}, we find that $\delta M^{(2)}_{R_x}$ is maximized when $\text{cos}(4 \theta)=-1$ so $\theta_{\text{opt}}=\pi/4$. 
In the cluster phase, we observe a noticeable dip in $\delta M^{(2)}_{R_x}$ at the location of maximal SRE, Eq.~(\ref{eq:gstar}), due to the MPS ground state already containing intrinsic nonstabilizerness in its construction. As $g \rightarrow 1$ we find that $\delta M^{(2)}_{R_x} \rightarrow0$ due to the iMPS state becoming the ground state of $H=\sum_{i}X_{i}$. 

Finally, the results for arbitrary single-qubit rotation  $U(\theta,\phi,\lambda)$ are remarkably similar to $x$-rotations. For example, $\delta M^{(2)}_{U}$ reaches a maximum of $\approx 0.59$ in the cluster phase, which occurs when the iMPS tensor becomes the $|T\rangle=(|0\rangle+e^{i\pi/4}|1\rangle)/\sqrt{2}$ state \cite{Leone2022}. Consistent with $\delta M^{(2)}_{R_{x}}$, there is a noticeable dip at the point of maximal SRE. In the product phase, we observe a noticeable drop as $g\rightarrow1$ to a constant value of $\delta M^{(2)}_{U} \approx0.41$ corresponding to the nonstabilizer state $|M\rangle=\cos(\beta/2)|0\rangle +e^{i\pi/4}\sin(\beta/2)|1\rangle$ with $\beta=\arccos(1/\sqrt{3})$. Thus, the injection of nonstabilizerness by single-site unitaries distinguishes the entangled SPT phase beyond its global SRE signatures.

\begin{figure}
    \centering
    \includegraphics[width=0.99\linewidth]{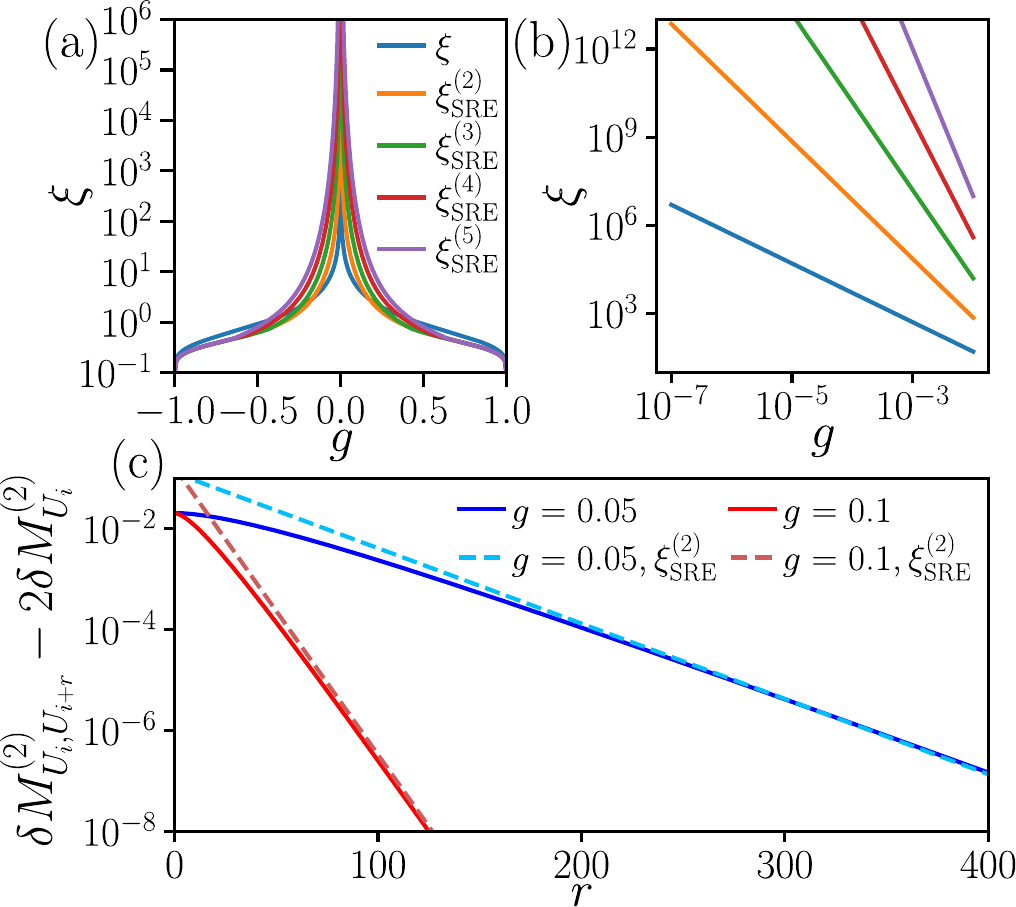}
    \caption{The SRE correlation length and response to perturbations of the MPS skeleton. (a): Correlation lengths of the standard transfer matrix $E$ and of the $n$th order SRE transfer matrix $\mathbb{E}$ for different numbers of replicas $n\leq5$. All correlation lengths diverge at the multicritical point, $g_c=0$. (b): Same data plotted on a log-log scale close to the critical point $g_c=0$. The ratio of the slopes of the blue and orange curves is $2$, consistent with Eq.~(\ref{eq:corrlengths_taylor}). Different correlation lengths generally diverge at different rates. (c): SRE correlations between two applied $\sqrt{T}$ gates for two values of $g$ close to $g_c=0$. The dashed lines are fits to the exponential decay with the SRE correlation length, which accurately describes the behavior of the correlators at large distances $r$ (see text).}
    \label{fig:Skeleton correlations}
\end{figure}

\subsection{Nonstabilizer correlations near criticality}

We now turn our attention to the SRE correlation length as a detector of the phase transition in the MPS skeleton. Using the analytically obtained eigenvalues of $E$ and $\mathbb{E}$, we obtain the following closed form expressions for the correlation lengths:
\begin{equation}\label{eq:corrlengths_full}
    \xi = \frac{-1}{\text{log}|\frac{1-|g|}{1+|g|}|}, \quad  \xi^{(2)}_{\mathrm{SRE}}= \frac{-1}{\text{log}|\frac{1-g^4}{1+14g^2 +g^4}|}. 
\end{equation}
Expanding these correlation length expressions around the critical point $g_c=0$:
\begin{equation}\label{eq:corrlengths_taylor}
    \xi %&=\frac{-1}{\text{log}|\frac{1-g}{1+g}|}\approx\frac{-1}{\text{log}(1-2g)} 
    \approx \frac{1}{2|g|}, \quad 
    \xi^{(2)}_{\mathrm{SRE}} %&= \frac{-1}{\text{log}|\frac{1-g^4}{1+14g^2 +g^4}|} \approx\frac{-1}{\text{log}(1-14g^2)}
    \approx\frac{1}{14g^2}.
\end{equation}
This shows that the standard and SRE correlation lengths can have parametrically different scaling at criticality, with $\xi$ diverging as $|g|^{-1}$, while $\xi^{(2)}_{\mathrm{SRE}}$ diverges as $|g|^{-2}$ at the MPS skeleton multicritical point. 

In Fig.~\ref{fig:Skeleton correlations}(a) we plot the correlation length $\xi^{(n)}_{\mathrm{SRE}}$ obtained by numerically diagonalizing the transfer matrix $\mathbb{E}$ for SRE orders $n\leq5$, alongside the standard correlation length $\xi$. Both types of correlation length diverge at the critical point, albeit with different rates and this rate moreover depends on the SRE order. This is demonstrated more transparently in Fig.~\ref{fig:Skeleton correlations}(b), which shows the correlation lengths in the vicinity of $g_c=0$ plotted on a log-log scale. While all curves are linear, their slope clearly varies with $n$, consistent with Eq.~\eqref{eq:corrlengths_taylor}.  

To confirm our prediction that nonstabilizer correlations between separated single qubit unitaries decay exponentially with respect to the SRE correlation length, we study $\delta M^{(2)}_{U_i,U_{i+r}} - 2\delta M^{(2)}_{U_i}$ to isolate the exponential decay. For the unitary, we choose the $U=\sqrt{T}=\text{diag}(1,e^{i\pi/8})$ gate, a single qubit gate that applies a $\pi/8$ phase rotation.
Figure~\ref{fig:Skeleton correlations}(c) shows the effect of separation distance $r$ on $\delta M^{(2)}_{U_{i},U_{i+r}} - 2\delta M^{(2)}_{U_i}$ for two values of $g$ close to the critical point. For both values of $g$, the correlation function approaches an exponential decay at large distances, which can be accurately fitted using the function $A \exp(-r/\xi_{\mathrm{SRE}}^{(2)})$, where $A$ is the fitting parameter and $\xi_{\mathrm{SRE}}^{(2)}$ is fixed by Eq.~\eqref{eq:corrlengths_full}. Therefore, the  SRE correlation length can detect nonstabilizer correlations between spatially-separated unitary operations. 

\begin{figure*}
    \centering
    \includegraphics[width=0.99\linewidth]{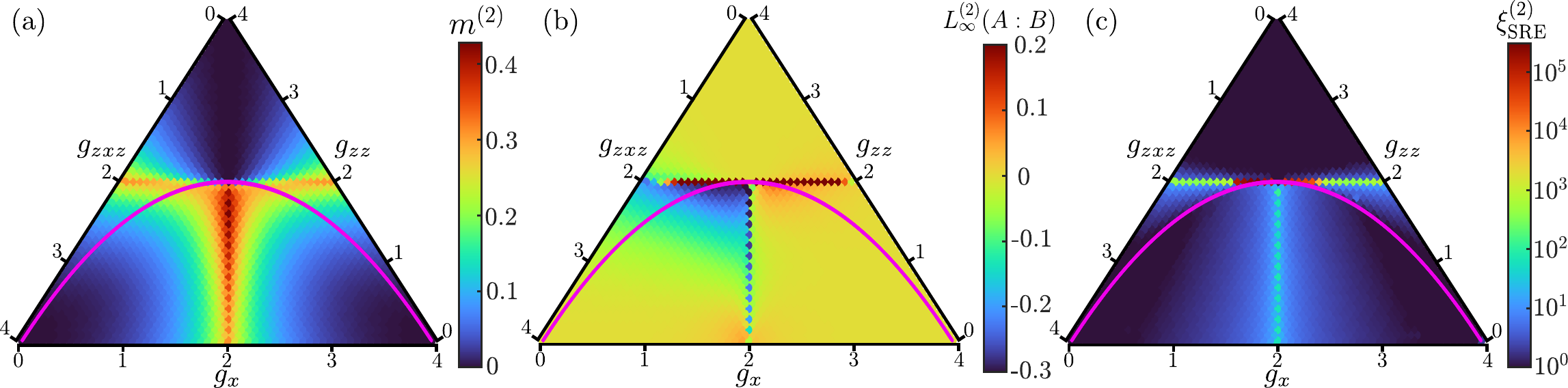}
    \caption{Phase diagram of nonstabilizerness in the cluster Ising model, Eq.~(\ref{eq:clustermodel}), in the thermodynamic limit. (a) The SRE density, $m^{(2)}$. (b) The mutual SRE, $L^{(2)}_\infty$. (c) The SRE correlation length, $\xi^{(2)}_{\mathrm{SRE}}$. All results are for iMPS with $\chi=50$ and $\chi_t=60$. The magenta line indicates the trajectory of the MPS skeleton, Eq.~(\ref{eq:MPS skeleton Hamiltonian}).
    } 
    \label{fig:cluster_ising_diagram}
\end{figure*}

\section{Nonstabilizerness in the cluster-Ising model}\label{sec:fullclusterIsing}

The MPS skeleton discussed previously traces a single-parameter trajectory through the phase diagram of the following spin-$\frac{1}{2}$ cluster-Ising model:
\begin{equation}\label{eq:clustermodel}
    H_{\text{CI}}=g_{zxz} \sum_i \sigma^{z}_{i}\sigma^{x}_{i+1}\sigma^{z}_{i+2} -g_{zz} \sum_i \sigma^{z}_{i}\sigma^{z}_{i+1} - g_{x}\sum_i \sigma^{x}_{i}.
\end{equation}
Here, the first three-body term is the cluster term that is added to the familiar Ising model in a transverse field (the last two terms).  While the model in Eq.~(\ref{eq:clustermodel}) is analytically solvable for general values of the couplings~\cite{Smacchia2011statmechclusterIsing}, the ground state is a simple MPS only along the special trajectory given by Eq.~(\ref{eq:MPS skeleton Hamiltonian}). 

 The Hamiltonian in Eq.~(\ref{eq:clustermodel}) is symmetric under global spin flip generated by $\prod_{i} \sigma^x_i$ and under time-reversal symmetry. Furthermore,  it is invariant under a Clifford control-Z unitary, $U_{CZ} \equiv \prod^N_{n=1} \mathrm{CZ}_{n}$, which transforms $\sigma^x_i \leftrightarrow \sigma^{z}_{i-1}\sigma^x_i\sigma^{z}_{i+1}$ and leaves $\sigma^{z}_{i}\sigma^{z}_{i+1}$ invariant. The phase diagram contains three phases: a symmetry-broken Ising phase, a trivial phase with a product ground state, and an SPT phase. These phases meet at the multicritical point that was studied in Sec.~\ref{sec:MPS_skeleton}, and here we investigate the behavior of nonstabilizerness across the entire phase diagram.  
 
\subsection{The phase diagram of nonstabilizerness}\label{subsec:cluster_ising_results}

For each choice of couplings $(g_{zxz}, g_x, g_{zz})$, we compute the ground state of the model in Eq.~(\ref{eq:clustermodel}) using the VUMPS algorithm \cite{Zauner2018VUMPS}. From the obtained iMPS matrices $A^{\sigma_i}$, we construct the corresponding Pauli-basis tensors $B^{\sigma_i}$, which are truncated to a computationally-manageable bond dimension $\chi_t$ in order to build the $n=2$ SRE replica transfer matrix of dimension $\chi_t^{4}$. 

In Fig.~\ref{fig:cluster_ising_diagram}(a) we show the resulting SRE density $m^{(2)}$ for $\chi = 50$ and $\chi_t = 60$. The diagram cleanly separates into three regions corresponding to the known phases of the model: the cluster-SPT phase (bottom left), the Ising paramagnetic phase (bottom right), and the ordered Ising phase (top). Because the unitary $U_{CZ}$ is Clifford and acts as a reflection about the vertical axis of the phase diagram, $m^{(2)}$ is symmetric about $g_x = 2$. Along the vertical critical line separating the SPT and Ising paramagnetic phases, $m^{(2)}$ exhibits a local maximum. This behavior contrasts sharply with that on the horizontal critical line, where $m^{(2)}$ interpolates from a local maximum at each endpoint to a pronounced local minimum at the central multicritical point.

Figure~\ref{fig:cluster_ising_diagram}(b) shows the mutual SRE of two arbitrarily large neighboring blocks $A$ and $B$, $L^{(2)}_{\infty}$, as defined in Eq.~\eqref{Eq:mutual SRE definition}. Since Rényi entropies are not subadditive, this quantity is not required to be positive. Indeed, while $L^{(2)}_{\infty}$ is positive throughout most of the phase diagram, it becomes negative across the SPT phase, particularly near the phase boundaries. Unlike $m^{(2)}$ in Fig.~\ref{fig:cluster_ising_diagram}(a), the mutual SRE is not vertically symmetric. Small regions of positive $L^{(2)}_{\infty}$ re-emerge near the phase boundary on the Ising–paramagnetic side.

Finally, the SRE correlation length, $\xi^{(2)}_{\mathrm{SRE}}$, shown in Fig.~\ref{fig:cluster_ising_diagram}(c), diverges along both critical lines, including at the multicritical point. Hence, its behavior is independent of either $m^{(2)}$ or $L^{(2)}_{\infty}$. This divergence highlights the usefulness of $\xi^{(2)}_{\mathrm{SRE}}$ as a diagnostic of critical nonstabilizerness and is discussed in more detail in the following subsection.

\subsection{Universal SRE scaling at criticality}\label{sec:critical_results}

\begin{figure*}
    \centering
    \includegraphics[width=0.99\linewidth]{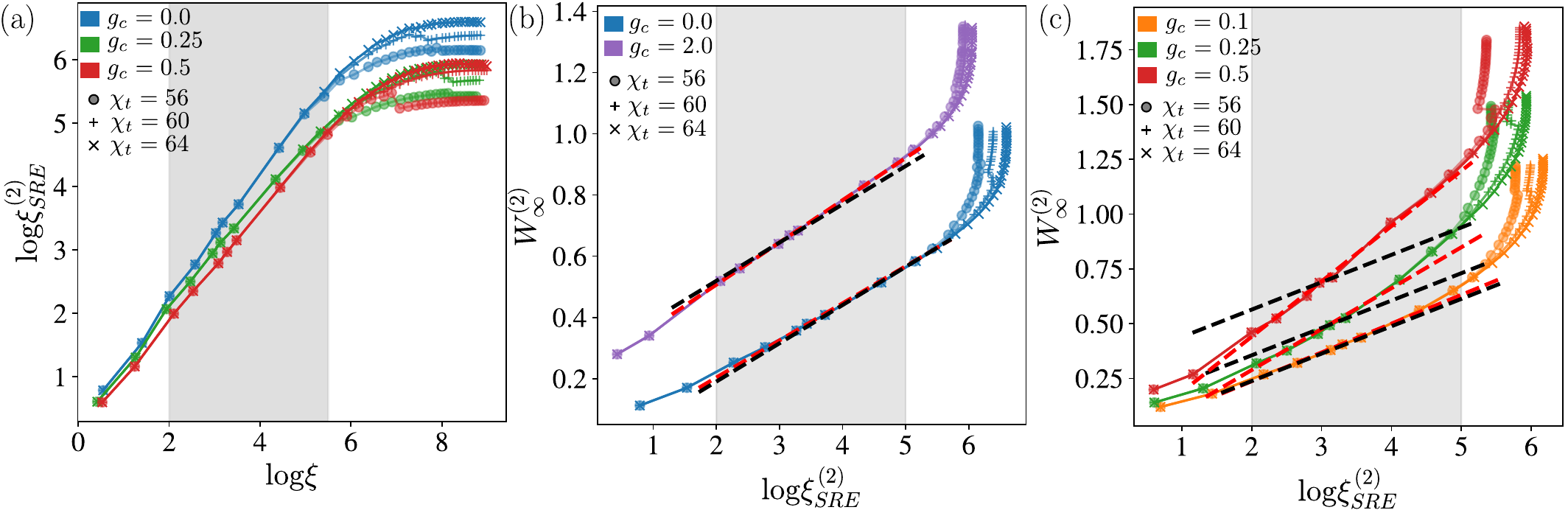}
    \caption{(a): The logarithm of the SRE correlation length $\log\xi_\mathrm{SRE}^{(2)}$, plotted as a function of the standard MPS correlation length $\log\xi$, at $g_c=0, 0.25,0.5$ along the horizontal cluster-Ising critical line for a few $\chi_t$ values indicated in the legend. The gray shaded area is the range where these two quantities are proportional to one another for the given values of $\chi_t$. (b): The mutual SRE density $W^{(2)}_{\infty}$ at the points $g_c=0$ and $g_c=2$ on the horizontal cluster-Ising critical line, plotted as a function of $\log \xi_\mathrm{SRE}^{(2)}$. The linear fit to the data is shown in red dashed lines, with the expected value $1/8$ shown in black, demonstrating good agreement. (c): Analogous results for the points $g_c=0.1,0.25, 0.5$ along the horizontal critical line, where the numerical data shows a visible deviation from the $1/8$ scaling. In (b) and (c), the noticeable upturn of $W^{(2)}_{\infty}$ and departure from linear dependence is due to the saturation of the SRE correlation length $\xi_\mathrm{SRE}^{(2)}$ in panel (a) for finite values of $\chi_t$.} 
    \label{fig:cluster_ising_critical}
\end{figure*}

The numerical results in Fig.~\ref{fig:cluster_ising_diagram} were obtained with bond dimensions $\chi=50$ and $\chi_t=60$, which are sufficient to produce well–converged results throughout most of the phase diagram. Even at criticality, the SRE density $m^{(2)}$ shows little sensitivity to either $\chi$ or $\chi_t$ and converges rapidly to a fixed value. In contrast, both the mutual SRE $L_\infty^{(n)}$ (and therefore also $W^{(n)}_{\infty}$) and the SRE correlation length $\xi^{(n)}_{\mathrm{SRE}}$ grow systematically as $\chi$ and $\chi_t$ are increased, reflecting the increasing accuracy with which the critical ground state is approximated as the effective length scale of the system increases.

Ref.~\cite{hoshino2025stabilizerrenyientropyconformal} proposed that, for a finite subregion of a periodic chain, the mutual SRE exhibits universal scaling, Eq.~(\ref{eq:BCFT2}), at a continuous transition in the Ising universality class. By contrast, the infinite, translation-invariant VUMPS solutions considered here directly computes $W^{(2)}_{\infty}$ for two semi-infinite subsystems sharing a single boundary. Although these subsystems are infinite, finite $\chi$ and $\chi_t$ induce a finite effective correlation length. By analogy with the finite-$\ell_c$ scaling in Eq.~(\ref{eq:BCFT2}), we therefore propose the asymptotic scaling form
\begin{equation}\label{eq:iMPSSREscaling}
W^{(n)}_{\infty}
= \frac{2\Delta_{2n}}{n-1}\log{\xi^{(n)}_{\mathrm{SRE}}} + b,
\end{equation}
which is expected to hold in the regime $\chi,\chi_t \gg 1$, up to a non-universal constant $b$. Note the factor-of-two difference compared to the periodic-boundary result in Eq.~(\ref{eq:BCFT2}), reflecting the single boundary between the two semi-infinite subsystems.

We next test Eq.~(\ref{eq:iMPSSREscaling}) numerically at different critical points in the cluster-Ising phase diagram, all belonging to the $\mathbb{Z}_2$ universality class where $\Delta_{2n}$ is expected to take the universal value $1/16$. In fact, for $n=2$ and the  $\mathbb{Z}_2$ universality class, $W^{(2)}_{\infty}$ and $I^{(2)}$ are predicted to obey the same scaling and therefore in $L^{(2)}_{\infty}$ the logarithmic scaling should cancel. Hence, we focus on $W_\infty^{(2)}$ and study its behavior along the horizontal critical line of the cluster-Ising model~(\ref{eq:clustermodel}), parametrized as $(g_c,2-g_c,2)$ with $g_c\in[0,2]$. Here $g_c=0$ corresponds to the Ising point studied in Refs.~\cite{hoshino2025stabilizerrenyientropyconformal,rajabpour2025stabilizershannonrenyiequivalenceexact}, while $g_c=1$ is the multicritical point analyzed in Sec.~\ref{sec:MPS_skeleton}. 

As we previously demonstrated for $g_c=1$ in Eq.~(\ref{eq:corrlengths_taylor}), the standard correlation length $\xi$ and the SRE correlation length $\xi_\mathrm{SRE}^{(2)}$ both diverge near criticality, albeit possibly with different exponents. Thus, we expect $\log \xi_\mathrm{SRE}^{(2)}$ to be proportional to $\log\xi$ for other values of $g_c$ along the critical line. We explicitly check this in  Fig.~\ref{fig:cluster_ising_critical}(a), which shows the proportionality holds in a relatively narrow window of $\xi$ that we are able to access with bond dimensions $\chi_t\leq 64$. Within this window, one may expect the iMPS results to be well-converged; outside of it, $\xi_\mathrm{SRE}^{(2)}$ saturates, meaning that $\chi_t$ is not large enough to fully capture the SRE correlations. The size of this window grows with $\chi_t$, although this comes with high computational cost.   

Despite these limitations, it is worth exploring the scaling of $W_\infty^{(2)}$. Figure~\ref{fig:cluster_ising_critical}(b) confirms that we approximately recover the expected scaling  at the endpoints of the critical line, $g_c=0$ and $g_c=2$: the red dashed lines show linear fits based on Eq.~(\ref{eq:iMPSSREscaling}) to iMPS data, yielding slopes in good agreement with the expected value $1/8$ (black dashed lines). Away from these endpoints, however, the approach to the asymptotic scaling regime is slower. For $g_c=0.1,\,0.25,\,0.5$, the extracted slopes are visibly larger than $1/8$, as seen in Fig.~\ref{fig:cluster_ising_critical}(c). While a residual dependence on $\chi_t$ cannot be excluded, the systematic bending of the curves suggests that the accessible range of $\xi$ is insufficient to exceed microscopic (ultraviolet) length scales, placing these data in a pre-asymptotic crossover regime. 

A few comments are in order. A similar overshoot of the slope for $g_c\in [0.1,0.5]$ (as well as $g_c\in [1.1, 1.5]$) is also observed in exact diagonalization studies of finite periodic systems, presented in Appendix~\ref{Appendix:ED}. This shows that the overshoot is not an artefact of the iMPS approach, but reflects the intrinsic difficulty of accessing the asymptotic scaling regime.  
Indeed, the standard properties of the underlying iMPS states, such as their entanglement entropy and correlation functions,  are well converged at these bond dimensions for all values of $g_c$.
Curiously, plotting $W^{(2)}_{\infty}$ as a function of  $\log \xi$ instead of $\log \xi_\mathrm{SRE}^{(2)}$ yields a linear behavior over a broader range, as shown in Appendix~\ref{Appendix:critical}; e.g., compare the data for the converged cases, $g_c=0,2$, in Fig.~\ref{fig:cluster_ising_critical}(b) with Fig.~\ref{fig:critical_xi}(a). We attribute this behavior to the different convergent properties of $c_1$ (which determines $W_\infty^{(2)}$), $\xi$ and $\xi_\mathrm{SRE}^{(2)}$: the first is determined by the \emph{leading} eigenvalue and eigenvector of $\mathbb{E}$, while the last two stem from the \emph{subleading} eigenvalue of $E$ and $\mathbb{E}$ respectively, with the latter expected to converge more slowly. Thus, our data appear to be in the regime where $c_1$ and $\xi$ are relatively well-converged, while $\xi_\mathrm{SRE}^{(2)}$ is not.
Nevertheless, by analyzing the dependence of $W_\infty^{(2)}$ on $\log\xi$ and extracting its slope, we find indications that the slope indeed approaches the universal value $1/8$ as we increase $\xi$, see Appendix~\ref{Appendix:critical} for further details. These results illustrate the challenges in accessing the universal scaling regime of the mutual SRE. For the model considered here, Eq.~(\ref{eq:clustermodel}), which can be mapped to free fermions, further insights could in principle be obtained using Gaussian state methods as in Refs.~\cite{collura2025quantummagicfermionicgaussian,rajabpour2025stabilizershannonrenyiequivalenceexact}, although our results suggest that the required system sizes may exceed thousands of sites.

\section{Conclusions}\label{sec:conclusions}

In this work, we introduced a spectral framework for characterizing nonstabilizerness based on the eigenspectrum of the SRE transfer matrix of iMPS states. By analyzing this spectrum, we showed that the SRE of finite subsystems embedded in an infinite system admits a natural decomposition into an extensive bulk contribution, a boundary term identified with the mutual SRE, and subleading corrections governed by an emergent SRE correlation length. This decomposition allows one to extract universal information contained in nonstabilizerness directly in the thermodynamic limit.

Applying this framework to the cluster-Ising model, we argued that the mutual SRE encodes universal information along critical lines, despite the nonuniversal behavior (at leading order) of the SRE density itself. Moreover, we showed that the SRE correlation length diverges at continuous phase transitions, providing a robust diagnostic of criticality even in situations where the SRE exhibits smooth or weakly nonanalytic behavior. Importantly, this correlation length is generally distinct from the conventional MPS correlation length and may diverge with a different critical exponent, highlighting that nonstabilizer correlations probe operator content beyond that captured by standard two-point functions.

A natural question posed by our results concerns the renormalization-group interpretation of finite bond dimension in the replicated theory underlying the SRE. While finite-$\chi$ scaling in conventional iMPS is well understood as an RG flow away from a critical fixed point in the physical CFT, our findings suggest that finite $\chi$ induces a more intricate flow in the $2n$-replica theory that controls nonstabilizerness. Developing a systematic CFT description of this flow would provide a deeper understanding of pre-asymptotic scaling regimes numerically observed in the mutual SRE and clarify how universal behavior emerges as $\chi$ is increased.

A key limitation of the present approach is the unfavorable scaling of the SRE transfer matrix with bond dimension, which makes numerically converged calculations increasingly costly at large $\chi$. One promising direction is to represent the boundary vectors themselves as finite $2n$-site MPS and to compute dominant eigenvalues using finite DMRG-type techniques, which may potentially enable access to larger effective bond dimensions. For models admitting a mapping to free fermions, such as the ones considered here, some nonstabilizer quantities (e.g., $M^{(2)}$) can be evaluated with complexity scaling linearly in system size~\cite{rajabpour2025stabilizershannonrenyiequivalenceexact}. Although such methods would need to be generalized to the mixed-state or mutual SRE studied here, they may provide an exact finite-size benchmark for how our transfer-matrix expressions approach the thermodynamic limit.
It would also be interesting to explore whether suitably generalized real-space RG methods~\cite{MartinDelgado1996} can access the replicated transfer-matrix spectrum underlying $\xi_{\mathrm{SRE}}^{(2)}$.

While the behavior of nonstabilizerness at the Ising critical point has now been studied in several works, many other critical regimes remain poorly understood. Our framework can be directly applied to other types of 1D critical behaviors, such as the XY critical line of the spin-$\frac{1}{2}$ XXZ chain or frustrated systems like the $J_1$-$J_2$ chain. Furthermore, measures of nonstabilizerness for higher-spin systems, such as the mana, could be formulated in terms of mana entropies with a structure closely analogous to the SRE. Beyond free-fermion models, natural targets for future investigations include other classes of tensor network skeletons, such as Onsager-integrable spin chains~\cite{camp2025matrixproductstateskeletonsonsagerintegrable} and two-dimensional Abelian string-net models~\cite{boesl2025skeletonisometrictensornetwork}. In this context, an important direction is the extension of these ideas to projected entangled-pair states (PEPS), which may shed light on how nonstabilizerness manifests in two-dimensional critical systems, for example in the Kitaev honeycomb lattice model or other types of two-dimensional lattice gauge theories. Beyond ground-state properties, our approach can be extended to dynamical settings described by time-evolving iMPS. The SRE correlation length introduced here offers a natural length scale for characterizing the spatial spreading of nonstabilizerness~\cite{maity2025localspreadingstabilizerrenyi,bejan2025magicspreadingunitaryclifford}, while the mutual SRE may encode universal dynamical scaling laws.

{\sl Note added.--}During the completion of this work, Ref.~\cite{nehra2025topologicalmagicresponsequantum} reported a complementary study of nonstabilizerness in ground states of Hamiltonians with SPT phases, for finite systems and uniformly doped with $\mathrm{T}$-gates. While our paper focuses on infinite systems and local perturbations, our results are consistent with theirs in comparable parameter regimes.

\section{Acknowledgments}

We thank M. A. Rajabpour, J. Ren, M. Bejan, T. Haug, G. Lami, and Y. Ashida for helpful discussions. Computational portions of this research were carried out on ARC4 and AIRE, part of the High-Performance Computing facilities at the University of Leeds. We acknowledge support by the Leverhulme Trust Research Leadership Award RL-2019-015 and EPSRC Grants EP/Z533634/1, UKRI1337. This research was supported in part by grant NSF PHY-2309135 to the Kavli Institute for Theoretical Physics (KITP). Z.P. acknowledges support by the Erwin Schr\"odinger International Institute for Mathematics and Physics. The data that supports the findings of this article is openly available at \cite{data_manuscript}.

\appendix

\section{Witnessing nonstabilizerness in iMPS}\label{app:witnessing magic}

As briefly discussed in Sec.~\ref{sec:overview}, the mixed-state SRE is considered a poor measure of mixed-state nonstabilizerness while still being a genuine monotone for mixed-state nonstabilizer resources. Recent works have pivoted to using efficient witnesses to classify the nonstabilizerness of mixed states. In this Appendix, we consider the $N$-qubit witness for a mixed state $\rho$~\cite{tarabunga2025efficientmutualmagicmagic} :
\begin{equation}\label{eq:witness}
    \mathcal{W}^{(n)}(\rho)=\frac{1}{1-n}\log{\sum_{P\in\mathcal{P}_N}\frac{|\text{Tr}(\rho P)|^{2n}}{2^N}} - \frac{1-2n}{1-n}S^{(n)}(\rho),
\end{equation}
which is related to the mixed-state nonstabilizerness via $\mathcal{W}^{(n)}\equiv \widetilde{M}^{(n)} - 2S^{(n)}$, thus allowing for easy calculation by nonstabilizerness replica tricks~\cite{Haug2023MPS,Banuls2024}. Importantly, $\mathcal{W}^{(n)}(\rho)$ is not a nonstabilizer monotone as it can be negative for some mixed states and may increase under applications of Clifford circuits. However, it can  identify mixed stabilizer states and provides a lower bound for other genuine monotones, like the log-free robustness of magic~\cite{Liu2022,Howard2017} and the mixed-state stabilizer fidelity~\cite{Bravyi2019simulationofquantum,Rubboli2024mixedstate}. Moreover, it is easy to express Eq.~\eqref{Eq:Magic Universal} and any other derivations in the main text using this witness by replacing the R\'enyi entropy with the rescaled R\'enyi entropy given in Eq.~\eqref{eq:witness}. For our iMPS calculations, we will focus on an associated witness density, $\mathcal{W}^{(n)}(\rho)/N$, which facilitates comparison to SREs in the thermodynamic limit.

\begin{figure}
    \centering
    \includegraphics[width=0.95\linewidth]{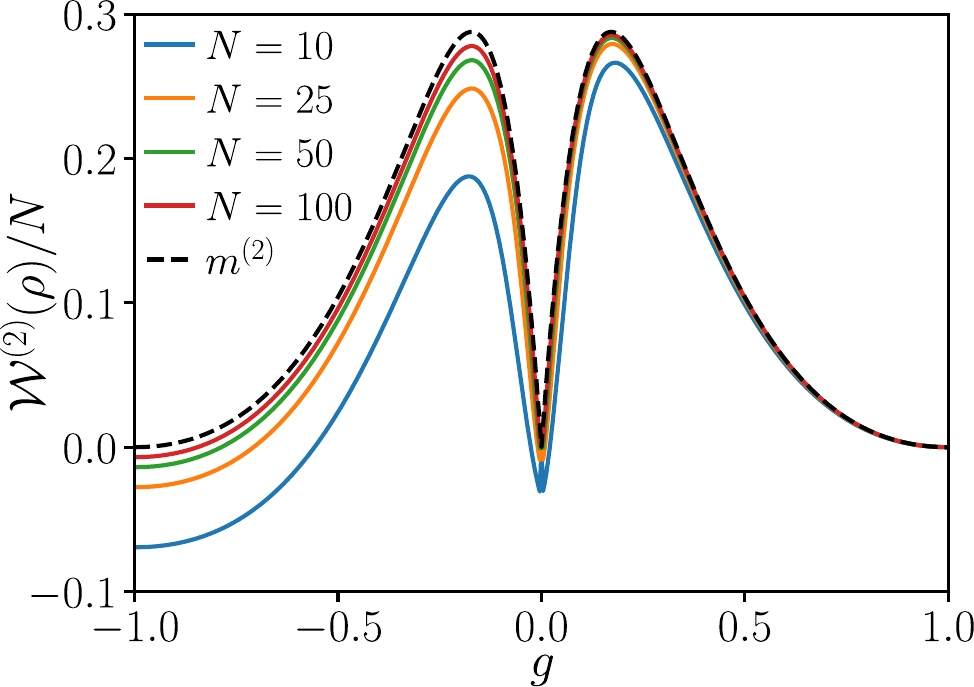}
    \caption{The witness of non-stabilizerness, Eq.~(\ref{eq:witness}) for $n=2$, for the skeleton iMPS and different subsystem sizes. $\mathcal{W}^{(2)}(\rho)/N$ was calculated via the replica trick given in Eq.~\eqref{eq:SRETM}. The black dashed line is the pure state SRE density $m^{(2)}$.}
    \label{fig:magic_witness}
\end{figure}

In Fig.~\ref{fig:magic_witness} we plot the witness density $\mathcal{W}^{(n)}(\rho)/N$ of the $\chi=2$ MPS skeleton, Eq.~(\ref{Eq:Skelton MPS}), for different subsystem sizes using direct computation, as done in Fig.~\ref{fig:Skeleton SRE Decay}. For comparison, we also include the pure state SRE density $m^{(2)}$, which is the pure-state nonstabilizerness of the MPS tensor in the thermodynamic limit. In the product state phase, we do not observe much change with subsystem size due to this phase mainly being dominated by the nonstabilizerness rather than entanglement, but we do see the limited amount of entanglement pushing $\mathcal{W}^{(n)}(\rho)$ to be more negative. In the SPT phase, the entanglement entropy has a larger effect and pushes $\mathcal{W}^{(n)}(\rho)$ to be negative, especially in the low nonstabilizer regime. By comparing with Fig.~\ref{fig:Skeleton SRE Decay}(a), we can see that the SRE density $m^{(2)}$ gives an upper bound to the witness but does not serve the same function for $\tilde{m}^{(2)}$. Finally we observe good convergence to $m^{(2)}$ with respect to subsystem size, indicating that our formalism can be easily extended to other measures of nonstabilizerness.

\section{SRE correlations between subsystems}\label{Appendix:SRE separated}

In Sec.~\ref{sec:magic_subsystems}, we obtained closed form relations for the nonstabilizerness of finite subsystems embedded in an infinite lattice, by considering the eigenspectrum of SRE replicated transfer matrices. While the nonstabilizerness of adjacent subsystems has been discussed in Refs.~\cite{Tarabunga2023,hoshino2025stabilizerrenyientropyconformal}, usual studies of correlations in MPS are undertaken for separated subsystems. In this appendix, we show that our formalism can be extended to non-adjacent subsystems, and we show that nonstabilizer correlations decay with respect to the standard correlation length of the \emph{unreplicated} iMPS~\cite{liu2025SRETIMPS}.

We begin by extending the standard transfer matrix $E$, Eq.~(\ref{eq:TMspectral}), to the replica space:
\begin{equation}\label{eq:repspectrum}
    E^{\otimes2n}=\sum_{i_{1,...,i_{2n}}}\left(\prod_{m=1}^{2n}\lambda_{i_{m}}\right)|R_{i_1}\otimes...\otimes R_{i_{2n}})(L_{i_1}\otimes...\otimes L_{i_{2n}}|.
\end{equation}
Since $\lambda_1=1$ (due to normalization), it is also the dominant eigenvalue of $E^{\otimes2n}$. Similarly, we can also identify the dominant eigenvectors as $(L_{1}\otimes...\otimes L_{1}|$ and $|R_1 \otimes...\otimes R_{1})$ which are just $(\mathbb{L}|, |\mathbb{R})$, respectively. Finally, the second dominant eigenvalue will also be given by $\lambda_2$ implying that the correlation length defined in Eq.~\eqref{Eq:correlation length} is the same as in the replicated space. For ease, we will denote the eigen-decomposition of the replicated transfer matrix as $E^{^{\otimes2n}}=\sum_{i}^{\chi^{4n}} \lambda^{(n)}_{i} |\mathbb{L}_{i})(\mathbb{R}_{i}|$.

To calculate the SRE of two subsystems $A$ and $B$ of equal length $N$ and separated by a distance $r$, we modify Eq.~\eqref{eq:expectation value} by inserting the eigendecomposition of $E^{\otimes2n}$:
\begin{equation} 
\begin{aligned}
    \langle \Lambda_{1:N}\Lambda_{N+r:2N+r}\rangle &= (\mathbb{L}|\mathbb{E}^N (E^{\otimes2n})^{r-1} \mathbb{E}^N|\mathbb{R}) \\
        &=\sum_{i}^{\chi^{4n}} (\lambda^{(n)}_{i})^{r-1} (\mathbb{L}|\mathbb{E}^N  |\mathbb{L}_{i})(\mathbb{R}_{i}| \mathbb{E}^N|\mathbb{R}).
\end{aligned}
\end{equation}
As in Sec.~\ref{subsec:SRE corrections}, we can more easily observe the impact of individual terms if we write out the summation explicitly:
\begin{multline}
    \langle \Lambda_{1:N}\Lambda_{N+r:2N+r}\rangle = (\mathbb{L}|\mathbb{E}^N  |\mathbb{L}_{1})(\mathbb{R}_{1}| \mathbb{E}^N|\mathbb{R}) + \\
    (\lambda^{(n)}_{2})^{r} (\mathbb{L}|\mathbb{E}^N  |\mathbb{L}_{2})(\mathbb{R}_{2}| \mathbb{E}^N|\mathbb{R}) + ...
\end{multline}
where we have set $\lambda^{(n)}_{1}=1$ due to normalization. We can easily see that the first term is just the expectation value of the two individual subsystems given in Eq.~\eqref{eq:expectation value} and for convenience we set $h_i=(\mathbb{L}|\mathbb{E}^N  |\mathbb{L}_{i})(\mathbb{R}_{i}| \mathbb{E}^N|\mathbb{R})$. 

The correlations between the subsystems $A$ and $B$ are given as:
\begin{equation}
    \langle \Lambda_{1:N}\Lambda_{N+r:2N+r}\rangle =\langle \Lambda_{1:N}\rangle^2 + (\lambda^{(n)}_{2})^{r-1}h_2 +...
\end{equation}
and hence the pure-state SRE as:
\begin{equation}
\begin{aligned}
        M^{(n)}(\rho_{AB}) &= \frac{\log\langle \Lambda_{1:N}\Lambda_{N+r:2N+r}\rangle}{1-n}\\
                &=\frac{\log[\langle \Lambda_{1:N}\rangle^2 +f(r)]}{1-n}, \;  f(r)\equiv \sum_{i=2}^{\chi^{4n}}(\lambda_{i}^{(n)})^{r-1}h_i.
\end{aligned}
\end{equation}
As in Sec.~\ref{subsec:SRE corrections}, since $f(r)\ll 1$, we use Taylor expansion to first order, leading to mixed-state SRE in our formalism:
\begin{eqnarray}\label{eq:mixed state magic separate}
    \nonumber \widetilde{M}^{(n)}(\rho_{AB}) &=& \frac{2\log(\langle \Lambda_{1:N}\rangle)}{1-n} -\frac{f(r)}{(1-n)\langle \Lambda_{1:N}\rangle^2} -S^{(2)}(\rho_{AB}) \\
    &+&\mathcal{O}\left(\frac{f(r)^2}{\langle \Lambda_{1:N}\rangle^2}\right).
\end{eqnarray}
If required, this can be further broken down using Eq.~\eqref{eq:pure magic expansion} to study the effect of subsystem size on the mixed-state SRE. However, since we are focusing on the behavior as a function of separation distance, we will refrain from decomposing the mixed state SRE further. Moreover, we can identify a leading term which is dominated by the SRE of the subsystems that are independent of each other and only depend on the size of each subsystem followed by lower-order terms which encode the SRE correlations between them. 

By keeping Eq.~\eqref{eq:mixed state magic separate} in this form, we can easily express the relationship between the mutual SRE and the separation of two finite subsystems. Recall the definition of the mutual SRE in Eq.~\eqref{eq:mutual magic}, %as $L^{(n)}(A:B)=\tilde{M}^{(n)}(\rho_A)+\tilde{M}^{(n)}(\rho_B)-\tilde{M}^{(n)}(\rho_{AB})$ 
which we know can be rewritten as $L^{(n)}=2\widetilde{M}^{(n)}(\rho_A)-\widetilde{M}^{(n)}(\rho_{AB})$ as the subsystems are of equal size. Using Eq.~\eqref{eq:pure magic expansion} and~\eqref{eq:mixed state magic separate}, the dominant terms cancel out and we can identify the mutual SRE:
\begin{equation}\label{eq:mutual SRE seperate cancelled}
    L^{(n)}(A:B)= \frac{f(r)}{(1-n)\langle \Lambda_{1:N}\rangle^2} - I^{(2)}(A:B).
\end{equation}

Finally, it is well-known that the dominant correlations in mutual information decay exponentially according to the correlation length $\xi$~\cite{Haag2023MI}, thus we can easily see that $f(r)$ also decays according to $\xi$. Ignoring any lower-order correlations, $f(r)\approx(\lambda^{(n)}_2)^r h_2 =e^{-r/\xi}h_2$ since $\lambda^{(n)}_2 = \lambda_2$. Therefore, since $\langle \Lambda_{1:N}\rangle$ is independent of $r$ and both $f(r)$ and $I^{(2)}(A:B)$ decay exponentially with respect to $\xi$ we infer that $L^{(n)}(A:B)$ should also decay exponentially, consistent with Ref.~\cite{liu2025SRETIMPS}.

\section{Higher order MPS skeletons}\label{Appendix:Higher order skeleton}

In Sec.~\ref{sec:fullclusterIsing} we introduced the cluster-Ising model and its $\chi=2$ MPS skeleton -- a minimal model of a topological quantum phase transition. This model is a member of a family of Hamiltonians that contain more complex transitions between SPT phases. Here we introduce the generalized cluster Hamiltonian and briefly review its connection to higher-order MPS skeletons via a Laurent polynomial encoding~\cite{Jones2021Skeletons}. We then use this encoding to verify our results from the main text using a $\chi=4$ MPS skeleton that exhibits phase transitions between a product phase and two distinct SPT phases.

We start from a free-fermion Hamiltonian~\cite{Altland1997SymmetryClass}:
\begin{equation}\label{eq:Hamiltonian majorana}
    H=\frac{1}{2}\sum_{n,\alpha}t_{\alpha}i\tilde{\gamma}_{n}\gamma_{n+\alpha},
\end{equation}
where $\gamma_{n} (\tilde{\gamma}_{n})$ are real (imaginary) Majorana operators, with real coefficients $t_{\alpha}$. It is easily seen that this Hamiltonian involves generalized cluster terms after performing the Jordan-Wigner transformation:
\begin{equation}
    i\tilde{\gamma_{n}}\gamma_{n+\alpha}=\Biggl\{ \begin{array}{ccc}
        -X_nZ_{n+1}...Z_{n+\alpha-1}X_{n+\alpha} & \text{if}&\alpha>0, \\
        Z_{n} & \text{if} &\alpha=0,\\
        -Y_{n+\alpha}Z_{n+\alpha+1}...Z_{n-1}Y_{n}& \text{if} &\alpha<0.
    \end{array}
\end{equation}
The couplings $t_\alpha$ of the Hamiltonian can be encoded into a Laurent polynomial~\cite{Verresen2018Topology,Jones2019Topology},
\begin{equation}
    f(z) = \sum_{\alpha}t_{\alpha}z^{\alpha},
\end{equation}
which allows us to extract physical properties of the model, including its single particle spectrum $\epsilon_k$, the correlation length $\xi$ and the winding number $\omega$. Furthermore, the ground state of Eq.~\eqref{eq:Hamiltonian majorana} can be represented as an MPS with finite bond dimension $\chi$ if 
\begin{equation}
 %f(z)=z^{p}g(z)^{2} \quad \text{and} \quad   g(z)=\sum_{k=0}^{d}s_{k}z^{k},
 f(z)=z^{p} \left( \sum_{k=0}^{d}s_{k}z^{k} \right)^{2}, 
\end{equation}
for some integers $p$ and $d$ and real coefficients $s_k$~\cite{Jones2021Skeletons}. As an example, the $\chi=2$ skeleton in Eq.~\eqref{Eq:Skelton MPS}, up to a gauge transformation, is generated by $d=1$, $p=0$ polynomial. 

\begin{figure*}
    \includegraphics[width=0.99\linewidth]{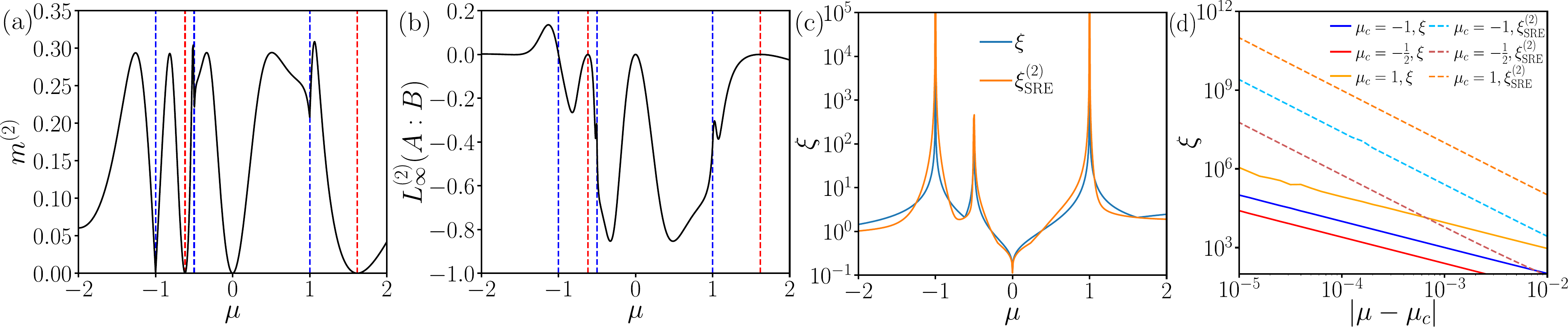}
    \caption{Nonstabilizerness of the $\chi=4$ MPS skeleton in Eq.~\eqref{Eq:Skelton chi4}. (a): The SRE density $m^{(2)}$ calculated via the dominant eigenvalue $\mu_1$ of the transfer matrix $\mathbb{E}$. (b): The  $n=2$ mutual SRE for two semi-infinite subsystems embedded in an infinite, translationally-invariant chain. The vertical blue and red dashed lines in panels (a)-(b) denote points where $|b_1|=1$ and $|b_2|=1$, respectively. (c): The correlation length and the $n=2$ SRE correlation length. (d): The same data as (c) plotted on a log-log scale close to the critical points $\mu_c=-1,-\frac{1}{2},1$ given as the blue, red and orange lines, respectively. The solid lines represent the standard correlation length, obtained from the transfer matrix $E$, while the dashed lines are the $n=2$ SRE correlation length, obtained from the SRE transfer matrix $\mathbb{E}$. Similar to the $\chi=2$ MPS skeleton in Fig.~\ref{fig:Skeleton correlations}, we find that the SRE correlation length $\xi_\mathrm{SRE}^{(2)}$ diverges faster than the ordinary correlation length $\xi$.}
    \label{fig:chi4_skeleton_plots}
\end{figure*}

To obtain a phase diagram with transitions to higher order SPT phases beyond the cluster-Ising model, we consider MPS skeletons that contains transitions between the $\omega=0,2,4$ SPT phases that are protected by the $\mathbb{Z}_2 \cross\mathbb{Z}_2^{T}$ symmetry that is generated by parity and complex conjugation operators. There exist three trajectories that can be expressed as a finite-$\chi$ MPS skeleton~\cite{Jones2021Skeletons}. For illustration, we consider the path described by the $d=2$, $p=0$ polynomial:
\begin{equation}\label{eq:chi4_skeleton_polynomial}
    f(z)=(z-\mu)^2 \left(z-\frac{\mu}{\mu+1}\right)^2,
\end{equation}
which describes the skeleton's trajectory through the 5-body generalized cluster model given by the Hamiltonian:
\begin{widetext}
\begin{eqnarray}
    \nonumber H &=& \frac{1}{2}\sum_n \frac{\mu^4}{(\mu+1)^2}Z_n +\left(\frac{2\mu^3}{\mu+1}+\frac{2\mu^3}{(\mu+1)^2}\right)X_n X_{n+1} - \left(\mu^2 +\frac{4\mu^2}{\mu+1}+\frac{\mu^2}{(\mu+1)^2}\right)X_n Z_{n+1}X_{n+2}\\ 
    &+& \left(\frac{2\mu}{\mu+1}+2\mu\right)X_n Z_{n+1}Z_{n+2}X_{n+3} - X_{n}Z_{n+1}Z_{n+2}Z_{n+3}X_{n+4}.
\end{eqnarray}
The ground state along this trajectory is exactly described by the $\chi=4$ MPS with spin matrices:
\begin{equation}\label{Eq:Skelton chi4}
    A^{\uparrow}=\begin{pmatrix}
        0&a_1&1&0\\
        a_2&0&0&-a_1a_2\\
        a_1&0&0&1\\
        0&-a_2&a_1a_2&0
    \end{pmatrix}, \;\;\;
    A^{\downarrow}=\begin{pmatrix}
        a_2&0&0&-a_1a_2\\
        0&a_1&1&0\\
        0&-a_2&a_1a_2&0\\
        a_1&0&0&1
    \end{pmatrix}, \;\; a_k=\frac{b_k}{1+\sqrt{1-b_k^2}}, b_1=-\frac{\mu(\mu+1)}{\mu^2 +\mu +1}, b_2=\frac{\mu+1}{\mu^2}.
\end{equation}
\end{widetext}
This model contains two sets of interesting points that correspond to $|b_1|=1$ or $|b_2|=1$. When $\mu\in\{-1,-\frac{1}{2},1\}$, then $|b_1|=1$, which represents the locations of the phase transitions that will be of main interest to us. However, when $\mu=(1\pm \sqrt{5})/2$ and $|b_2|=1$, we obtain the cluster state which allows us to draw parallels between the skeleton in Eq.~\eqref{Eq:Skelton chi4} and the skeleton we previously studied in Eq.~\eqref{Eq:Skelton MPS}.

In Fig.~\ref{fig:chi4_skeleton_plots}(a) we plot the $n=2$ SRE of the $\chi=4$ MPS skeleton obtained from the dominant eigenvalue of its replica transfer matrix. First, for both values of $\mu$ where $|b_2|=1$, $m^{(2)}=0$ which is consistent with this MPS representing the cluster state---a well-known stabilizer state. Interestingly, when $|b_1|=1$, we only see $m^{(2)}=0$ for $\mu=-1$ due to it being at the phase transition between the $\omega=0$ and $\omega=2$ SPT phases, which we know is the GHZ state and hence a stabilizer state. For $\mu=-1/2,1$ which represents the $\omega=2 \to \omega=4$ phase transition (and vice versa), we observe $m^{(2)}\approx 0.20$. The generating Laurent polynomials for the two phase transitions are $f(z)=z(z+1/2)^2$ and $f(z)=-z(z-1/2)^2$ for $\mu=-1/2$ and $\mu=1$, respectively, which are just the Laurent polynomials of the $\chi=2$ skeleton for $g=1/3$ and $g=3$. During the construction of the MPS the extra $z$ terms in the polynomials represent the application of a `SPT Entangler' \cite{Jones2021Skeletons}, which is a stabilizer operation and does not affect the SRE. Therefore, by Eq.~\eqref{eq:skeleton peak}, we see that $m^{(2)}=-\text{log}(13/16) \approx0.20$ implies that these topological phase transitions can also be described by considering skeletons of lower order. Finally, located at $\mu=0$ is another stabilizer state as it is the ground state of the parent Hamiltonian $H=-1/2 \sum_{n} X_{n}Z_{n+1}Z_{n+2}Z_{n+3}X_{n+4}$, which is a 5-body stabilizer code.

In Fig.~\ref{fig:chi4_skeleton_plots}(b) we plot the mutual SRE obtained from the overlap between the dominant eigenvectors of the SRE transfer matrix $\mathbb{E}$ and the replicated dominant eigenvectors of the standard transfer matrix $E$ for the $\chi=4$ skeleton. For the $\omega=0 \rightarrow\omega=2$ transitions we observe behavior consistent with Fig.~\ref{fig:Skeleton SRE Decay}(b). In the $\omega=4$ SPT phase, $L^{(2)}_\infty \approx-0.85$, indicating that entanglement heavily dominates the SRE term, which is due to the entanglement in that region saturating the upper bound of $S^{(2)}(\rho)=2\log 2$. It is also notable that for the $\omega=2 \rightarrow\omega=4$ phase transition, we obtain a non-zero mutual SRE. For this transition, $L^{(2)}_\infty \approx-0.38$ and can be explained similarly by considering that the generating Laurent polynomial is equal to that of the $\chi=2$ MPS, with the only difference being the larger entanglement entropy at the phase transition.

In Fig.~\ref{fig:chi4_skeleton_plots}(c) we plot the correlation lengths obtained from the standard transfer matrix $E$ and from the $n=2$ SRE transfer matrix $\mathbb{E}$. We find that $\xi^{(2)}_{\mathrm{SRE}}$ clearly identifies the three phase transitions alongside the standard correlation length. We also observe consistent behavior in the correlation lengths compared to Fig.~\ref{fig:Skeleton correlations}(a) with $\xi^{(2)}_{\mathrm{SRE}}$ diverging much faster compared to $\xi$. This is further demonstrated by Fig.~\ref{fig:chi4_skeleton_plots}(d) where the correlation lengths are plotted on a log-log scale close to the critical points $\mu_c=-1,-\frac{1}{2},1$. As observed for the $\chi=2$ MPS skeleton in Sec.~\ref{sec:MPS_skeleton}, all curves are linear with the ratios between the slopes $\xi$ and $\xi^{(2)}_{\mathrm{SRE}}$ being approximately equal to 2 for all three critical points. 

\begin{figure}
    \centering
    \includegraphics[width=0.99\linewidth]{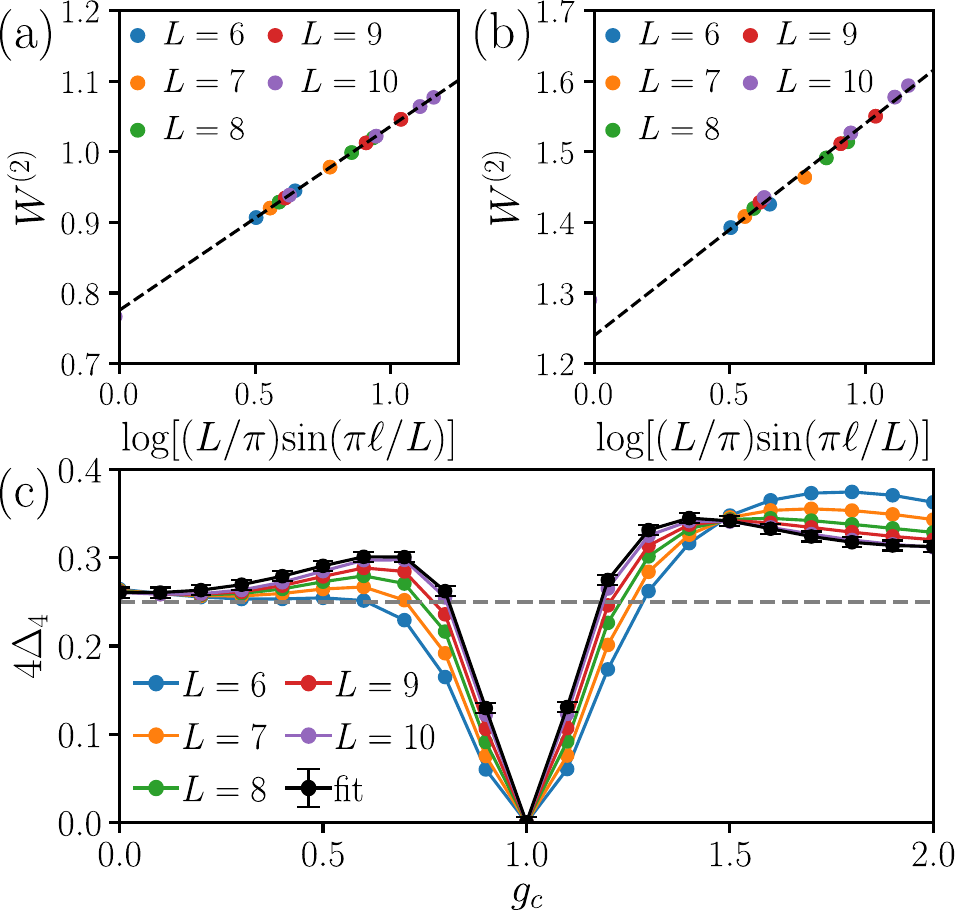}
    \caption{(a)-(b): The $n=2$ mutual SRE for multiple system sizes $L$ with PBCs and all admissible subsystem sizes $\ell$. Panel (a) is for $g_c=0$, while panel (b) is for $g_c=0.7$. The black dashed lines are the linear fits to extract the slope $4\Delta_{4}$ according to Eq.~\eqref{eq:BCFT2}. (c): The extracted $4\Delta_{4}$ across the critical line for different system sizes $L$, along with their extrapolation to the thermodynamic limit (black line). The gray dashed line is the exact asymptotic value of $4\Delta_{4}=1/4$~\cite{hoshino2025stabilizerrenyientropyconformal,rajabpour2025stabilizershannonrenyiequivalenceexact}.}
    \label{fig:Critical_line_ED}
\end{figure}

\section{Exact diagonalization study of finite systems with periodic boundary conditions}\label{Appendix:ED}

In Sec.~\ref{sec:critical_results} we probed the universal scaling of mutual SRE in the thermodynamic limit along the $\mathbb{Z}_2$ critical line of the model in Eq.~\eqref{eq:clustermodel}. At the Ising point ($g_c=0$) and the cluster Ising point ($g_c=2$), the results were found to be consistent with the formula~(\ref{eq:BCFT2}) in the limit $\chi\to\infty$~\cite{hoshino2025stabilizerrenyientropyconformal,rajabpour2025stabilizershannonrenyiequivalenceexact}. However, in the vicinity of the multicritical point $g_c=1$, the iMPS result based on Eq.~\eqref{eq:iMPSSREscaling} overshoots the exact asymptotic value. Here we attempt to reconcile these results by studying finite systems with periodic boundary conditions (PBCs) using exact diagonalization. In computing the SRE, we perform a brute force numerical evaluation of all $\mathbb{Z}_2$-preserving Pauli strings~\cite{Tarabunga2023}.

In Fig.~\ref{fig:Critical_line_ED}(a)-(b) we show the $n=2$ pure state mutual SRE as a function of subsystem size for the critical Ising model, i.e., $g_c=0$  in  Eq.~\eqref{eq:clustermodel}. We collect the data for system sizes $L=6-10$ and fit them according to Eq.~\eqref{eq:BCFT2} to extract $4\Delta_{4}$, which is shown by the black dashed line. We obtain  excellent agreement between the extracted value $4\Delta_{4}\approx 0.26$ and the predicted  $4\Delta_{4}=1/4$ for $g_c=0$ [Fig.~\ref{fig:Critical_line_ED}(a)]. On the other hand, Fig.~\ref{fig:Critical_line_ED}(b) repeats the same analysis closer to the multicritical point at $g_c=0.7$,  where the linear scaling still appears to hold, but the extracted slope $4\Delta_{4}\approx0.30$ is much further away from the predicted scaling.  

The mutual SRE dependence on $g_c$ along the critical line is summarized in Fig.\ref{fig:Critical_line_ED}(c), where the black line is the fit to $L\to\infty$. As noted previously, we obtain good agreement near the Ising point $g_c=0$. Moreover, the ground state at the multicritical point $g_c=1$ is the GHZ state, hence the mutual SRE should be zero, as indeed reproduced by the numerical data. However, in between these points, we observe significant deviations from the expected value of $4\Delta_{4}$. For example, at the cluster-Ising point $g_c=2$, the extrapolated slope is $4\Delta_{4}\approx0.31$, although the data trend (decrease with $L$) is consistent with convergence towards the asymptotic slope $1/4$ in larger systems. On the other hand, the largest deviation from the exact scaling is at $g_c=1.4$, where the extracted coefficient is  $4\Delta_{4}\approx0.34$. Moreover, in this case (similar to $g_c\sim 0.7$) the data drifts \emph{away} from the BCFT prediction with increasing $L$. This suggests that there is an emergent (ultraviolet) length scale around $g_c\approx 1$ which is larger than the values of $L$ in Fig.~\ref{fig:Critical_line_ED}, preventing the observation of universal scaling. We note that our iMPS results in the main text are qualitatively consistent with exact results in Fig.~\ref{fig:Critical_line_ED}, in particular the iMPS also overshoots the predicted slope in the regime $0.5 \lesssim g_c < 1$. 

\section{Extracting universal critical behavior of nonstabilizerness}\label{Appendix:critical}
\begin{figure*}
    \centering
    \includegraphics[width=0.99\linewidth]{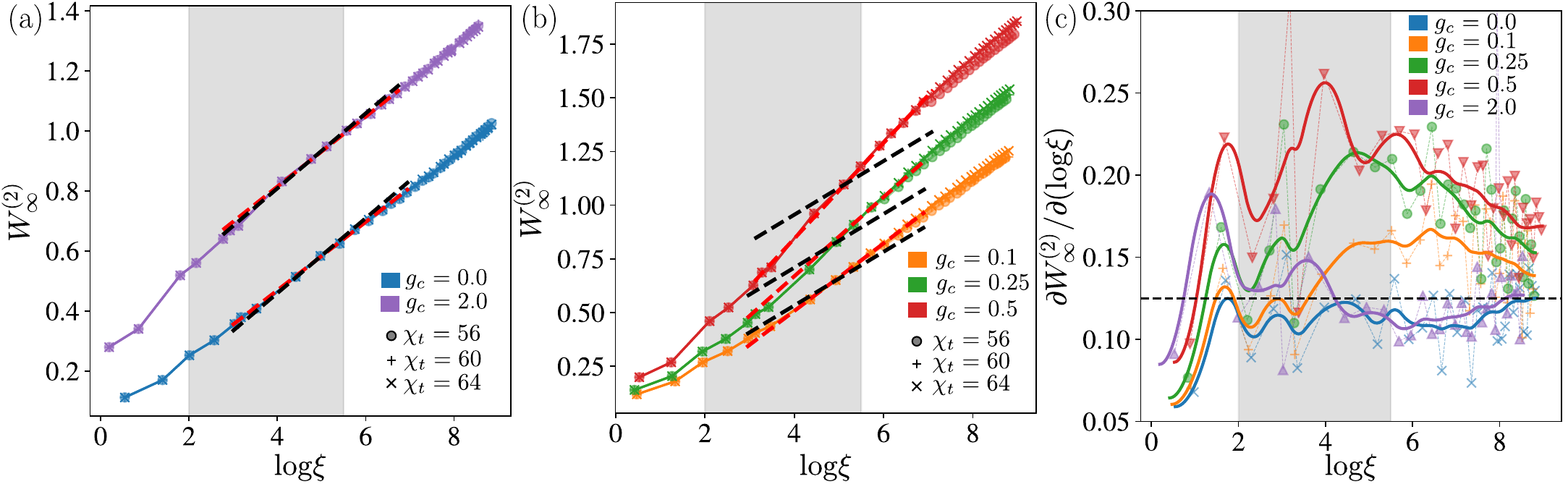}
    \caption{(a): The mutual SRE density $W^{(2)}_{\infty}$ at the points $g_c=0$ and $g_c=2$ on the horizontal cluster-Ising critical line, plotted as a function of $\log \xi$. The linear fit to the data is shown in red dashed lines, with the expected value $1/8$ shown in black, demonstrating good agreement. (b): Analogous results for the points $g_c=0.1,0.25, 0.5$ along the horizontal critical line, where the numerical data shows a visible deviation from the $1/8$ scaling. Nevertheless, we also observe the bending of curves at large values of $\xi$, indicating that the data may be in a pre-asymptotic regime. (c): The instantaneous numerical estimate of the gradient of $W^{(2)}_{\infty}$ against $\log \xi$ using the results presented in (b)-(c) for $\chi_t=64$. Despite large fluctuations in the data, all values of $g_c$ are consistent with an approach to the $1/8$ scaling at large $\xi$. The solid lines are a spline interpolation of the data points and serve as a guide to the eye.  All results were obtained for iMPS with $\chi\in [2,120]$ and $\chi_t=56,60,64$.} 
    \label{fig:critical_xi}
\end{figure*}

In the main text, we proposed that $W^{(2)}_\infty$ satisfies the scaling form of Eq.~\eqref{eq:iMPSSREscaling}, with $\xi^{(2)}_{\mathrm{SRE}}$ effectively playing the role of system size. However, in practice, obtaining well-converged $\xi^{(2)}_{\mathrm{SRE}}$ is computationally demanding due to its pronounced sensitivity to $\chi_t$, as seen in Fig.~\ref{fig:cluster_ising_critical} for values $0< g_c < 1$. Since $\xi^{(2)}_{\mathrm{SRE}}$ is extracted from the subleading eigenvalues of the $n=2$ SRE transfer matrix, it is expected to be substantially more sensitive to $\chi_t$ compared to $W^{(2)}_\infty$, as the latter depends on the dominant eigenvector of the same transfer matrix.

Motivated by this observation, here we make an attempt at interpreting Fig.~\ref{fig:cluster_ising_critical} using a modified scaling form 
\begin{equation}\label{eq:iMPSorigscaling}
W^{(n)}_{\infty}
= \frac{2\Delta_{2n}}{n-1}\log{\xi} + b,
\end{equation}
where $\xi^{(n)}_{\mathrm{SRE}}$ is replaced by the standard MPS correlation length $\xi$. Figure~\ref{fig:critical_xi}(a) demonstrates that this formula describes well the cases $g_c=0$ and $g_c=2$, where linear dependence $W_\infty^{(2)}\propto \log \xi$ is essentially observed over the full data range, i.e., not just in the regime where the two length scales are proportional to each other, $\xi^{(2)}_\mathrm{SRE} \sim \xi^p$, but also where $\xi^{(2)}_\mathrm{SRE}$ appeared to saturate in Fig.~\ref{fig:cluster_ising_critical}(a).  

Away from the endpoints, the dependence $W_\infty^{(2)}\propto \log \xi$ still approximately holds over a relatively broad range, as seen in Fig.~\ref{fig:critical_xi}(b). Nevertheless, there are clear systematic deviations from the universal $1/8$ slope, shown by dashed black lines. The deviations become more pronounced as $g_c$ approaches $0.5$, consistent with our analysis in Fig.~\ref{fig:cluster_ising_critical}(c). The deviations are explored more systematically in Fig.~\ref{fig:critical_xi}(c), where we compute the instantaneous slope of $W^{(2)}_{\infty}$ with respect to $\log \xi$. Although the numerical derivative is noisy, spline interpolation reveals a clear trend. For $g_c=0$ and $g_c=2$, the slope remains close to $1/8$ and approaches it more closely as $\xi$ increases. For intermediate values of $g_c$, the slope consistently overshoots $1/8$, but exhibits a systematic downward drift toward this value with increasing $\xi$, consistent with a crossover toward the expected universal Ising scaling. Thus, the observed deviations likely reflect a pre-asymptotic regime, with the true universal scaling emerging only at larger $\xi$. 
However, one must keep in mind that the SRE correlation length $\xi^{(2)}_{\mathrm{SRE}}$ is not fully converged for the largest values of $\chi$ in Fig.~\ref{fig:critical_xi}, leaving some uncertainty as to whether the behavior of $W^{(2)}_\infty$ in this regime can be regarded as quantitatively reliable.

\bibliography{references}

@article{smith2024Rydbergatoms,
  title = {Nonstabilizerness in kinetically constrained {R}ydberg atom arrays},
  author = {Smith, Ryan and Papi\ifmmode \acute{c}\else \'{c}\fi{}, Zlatko and Hallam, Andrew},
  journal = {Phys. Rev. B},
  volume = {111},
  issue = {24},
  pages = {245148},
  numpages = {11},
  year = {2025},
  month = {Jun},
  publisher = {American Physical Society},
  doi = {10.1103/jz4d-vdhj},
  url = {https://link.aps.org/doi/10.1103/jz4d-vdhj}
}

@article{MartinDelgado1996,
author = {MART\'{I}N-DELGADO, MIGUEL A. and SIERRA, GERM\'{A}N},
title = {ANALYTIC FORMULATIONS OF THE DENSITY MATRIX RENORMALIZATION GROUP},
journal = {International Journal of Modern Physics A},
volume = {11},
number = {17},
pages = {3145-3174},
year = {1996},
doi = {10.1142/S0217751X96001516},
URL = { 
        https://doi.org/10.1142/S0217751X96001516
},
eprint = { }
}

@misc{ahmad2025experimentaldemonstrationnonlocalmagic,
      title={Experimental demonstration of non-local magic in a superconducting quantum processor}, 
      author={Halima Giovanna Ahmad and Gianluca Esposito and Viviana Stasino and Jovan Odavic and Carlo Cosenza and Alessandro Sarno and Pasquale Mastrovito and Michele Viscardi and Stefano Cusumano and Francesco Tafuri and Davide Massarotti and Alioscia Hamma},
      year={2025},
      eprint={2511.15576},
      archivePrefix={arXiv},
      primaryClass={quant-ph},
      url={https://arxiv.org/abs/2511.15576}, 
}

@article{Cepollaro2025,
  title = {Harvesting stabilizer entropy and nonlocality from a quantum field},
  author = {Cepollaro, Simone and Cusumano, Stefano and Hamma, Alioscia and Lo Giudice, Giorgio and Odavi\ifmmode \acute{c}\else \'{c}\fi{}, Jovan},
  journal = {Phys. Rev. D},
  volume = {112},
  issue = {10},
  pages = {105012},
  numpages = {17},
  year = {2025},
  month = {Nov},
  publisher = {American Physical Society},
  doi = {10.1103/4brj-cl26},
  url = {https://link.aps.org/doi/10.1103/4brj-cl26}
}

@article{Qian2025,
  title = {Quantum nonlocal nonstabilizerness},
  author = {Qian, Dongheng and Wang, Jing},
  journal = {Phys. Rev. A},
  volume = {111},
  issue = {5},
  pages = {052443},
  numpages = {9},
  year = {2025},
  month = {May},
  publisher = {American Physical Society},
  doi = {10.1103/PhysRevA.111.052443},
  url = {https://link.aps.org/doi/10.1103/PhysRevA.111.052443}
}

@article{Haug2023MPS,
  title = {Quantifying nonstabilizerness of matrix product states},
  author = {Haug, Tobias and Piroli, Lorenzo},
  journal = {Phys. Rev. B},
  volume = {107},
  issue = {3},
  pages = {035148},
  numpages = {10},
  year = {2023},
  month = {Jan},
  publisher = {American Physical Society},
  doi = {10.1103/PhysRevB.107.035148},
  url = {https://link.aps.org/doi/10.1103/PhysRevB.107.035148}
}

@article{Howard2017,
  title = {Application of a Resource Theory for Magic States to Fault-Tolerant Quantum Computing},
  author = {Howard, Mark and Campbell, Earl},
  journal = {Phys. Rev. Lett.},
  volume = {118},
  issue = {9},
  pages = {090501},
  numpages = {6},
  year = {2017},
  month = {Mar},
  publisher = {American Physical Society},
  doi = {10.1103/PhysRevLett.118.090501},
  url = {https://link.aps.org/doi/10.1103/PhysRevLett.118.090501}
}

@article{Heinrich2019robustnessofmagic,
  doi = {10.22331/q-2019-04-08-132},
  url = {https://doi.org/10.22331/q-2019-04-08-132},
  title = {Robustness of {M}agic and {S}ymmetries of the {S}tabiliser {P}olytope},
  author = {Heinrich, Markus and Gross, David},
  journal = {{Quantum}},
  issn = {2521-327X},
  publisher = {{Verein zur F{\"{o}}rderung des Open Access Publizierens in den Quantenwissenschaften}},
  volume = {3},
  pages = {132},
  month = apr,
  year = {2019}
}

@article{Banuls2024,
  title = {Nonstabilizerness via Matrix Product States in the Pauli Basis},
  author = {Tarabunga, Poetri Sonya and Tirrito, Emanuele and Ba\~nuls, Mari Carmen and Dalmonte, Marcello},
  journal = {Phys. Rev. Lett.},
  volume = {133},
  issue = {1},
  pages = {010601},
  numpages = {7},
  year = {2024},
  month = {Jul},
  publisher = {American Physical Society},
  doi = {10.1103/PhysRevLett.133.010601},
  url = {https://link.aps.org/doi/10.1103/PhysRevLett.133.010601}
}

@article{lami2023quantum,
  title = {Nonstabilizerness via Perfect Pauli Sampling of Matrix Product States},
  author = {Lami, Guglielmo and Collura, Mario},
  journal = {Phys. Rev. Lett.},
  volume = {131},
  issue = {18},
  pages = {180401},
  numpages = {6},
  year = {2023},
  month = {Oct},
  publisher = {American Physical Society},
  doi = {10.1103/PhysRevLett.131.180401},
  url = {https://link.aps.org/doi/10.1103/PhysRevLett.131.180401}
}

@article{Sarkar_2020,
doi = {10.1088/1367-2630/aba919},
url = {https://dx.doi.org/10.1088/1367-2630/aba919},
year = {2020},
month = {aug},
publisher = {IOP Publishing},
volume = {22},
number = {8},
pages = {083077},
author = {S Sarkar and C Mukhopadhyay and A Bayat},
title = {Characterization of an operational quantum resource in a critical many-body system},
journal = {New Journal of Physics},
}

@Article{Oliviero2022,
author={Oliviero, Salvatore F. E.
and Leone, Lorenzo
and Hamma, Alioscia
and Lloyd, Seth},
title={Measuring magic on a quantum processor},
journal={npj Quantum Information},
year={2022},
month={Dec},
day={19},
volume={8},
number={1},
pages={148},
issn={2056-6387},
doi={10.1038/s41534-022-00666-5},
url={https://doi.org/10.1038/s41534-022-00666-5}
}

@Article{niroula2024phase,
author={Niroula, Pradeep
and White, Christopher David
and Wang, Qingfeng
and Johri, Sonika
and Zhu, Daiwei
and Monroe, Christopher
and Noel, Crystal
and Gullans, Michael J.},
title={Phase transition in magic with random quantum circuits},
journal={Nature Physics},
year={2024},
month={Nov},
day={01},
volume={20},
number={11},
pages={1786-1792},
issn={1745-2481},
doi={10.1038/s41567-024-02637-3},
url={https://doi.org/10.1038/s41567-024-02637-3}
}

@article{Leone2022Ising,
  title = {Magic-state resource theory for the ground state of the transverse-field {I}sing model},
  author = {Oliviero, Salvatore F. E. and Leone, Lorenzo and Hamma, Alioscia},
  journal = {Phys. Rev. A},
  volume = {106},
  issue = {4},
  pages = {042426},
  numpages = {6},
  year = {2022},
  month = {Oct},
  publisher = {American Physical Society},
  doi = {10.1103/PhysRevA.106.042426},
  url = {https://link.aps.org/doi/10.1103/PhysRevA.106.042426}
}

@Article{viscardi2025interplayentanglementstructuresstabilizer,
	title={{Interplay of entanglement structures and stabilizer entropy in spin models}},
	author={Michele Viscardi and Marcello Dalmonte and Alioscia Hamma and Emanuele Tirrito},
	journal={SciPost Phys. Core},
	volume={9},
	pages={012},
	year={2026},
	publisher={SciPost},
	doi={10.21468/SciPostPhysCore.9.1.012},
	url={https://scipost.org/10.21468/SciPostPhysCore.9.1.012},
}

@article{Wolf2006Skeleton,
  title = {Quantum Phase Transitions in Matrix Product Systems},
  author = {Wolf, Michael M. and Ortiz, Gerardo and Verstraete, Frank and Cirac, J. Ignacio},
  journal = {Phys. Rev. Lett.},
  volume = {97},
  issue = {11},
  pages = {110403},
  numpages = {4},
  year = {2006},
  month = {Sep},
  publisher = {American Physical Society},
  doi = {10.1103/PhysRevLett.97.110403},
  url = {https://link.aps.org/doi/10.1103/PhysRevLett.97.110403}
}

@article{Smith2022Ctopologicalphase,
  title = {Crossing a topological phase transition with a quantum computer},
  author = {Smith, Adam and Jobst, Bernhard and Green, Andrew G. and Pollmann, Frank},
  journal = {Phys. Rev. Res.},
  volume = {4},
  issue = {2},
  pages = {L022020},
  numpages = {8},
  year = {2022},
  month = {Apr},
  publisher = {American Physical Society},
  doi = {10.1103/PhysRevResearch.4.L022020},
  url = {https://link.aps.org/doi/10.1103/PhysRevResearch.4.L022020}
}

@article{Jones2021Skeletons,
  title = {Skeleton of matrix-product-state-solvable models connecting topological phases of matter},
  author = {Jones, Nick G. and Bibo, Julian and Jobst, Bernhard and Pollmann, Frank and Smith, Adam and Verresen, Ruben},
  journal = {Phys. Rev. Res.},
  volume = {3},
  issue = {3},
  pages = {033265},
  numpages = {26},
  year = {2021},
  month = {Sep},
  publisher = {American Physical Society},
  doi = {10.1103/PhysRevResearch.3.033265},
  url = {https://link.aps.org/doi/10.1103/PhysRevResearch.3.033265}
}

@article{Liu2025Ising,
  title = {Nonequilibrium quantum Monte Carlo algorithm for stabilizer {R}\'enyi entropy in spin systems},
  author = {Liu, Zejun and Clark, Bryan K.},
  journal = {Phys. Rev. B},
  volume = {111},
  issue = {8},
  pages = {085144},
  numpages = {14},
  year = {2025},
  month = {Feb},
  publisher = {American Physical Society},
  doi = {10.1103/PhysRevB.111.085144},
  url = {https://link.aps.org/doi/10.1103/PhysRevB.111.085144}
}

@article{SCHOLLWOCK201196,
title = {The density-matrix renormalization group in the age of matrix product states},
journal = {Annals of Physics},
volume = {326},
number = {1},
pages = {96-192},
year = {2011},
note = {January 2011 Special Issue},
issn = {0003-4916},
doi = {https://doi.org/10.1016/j.aop.2010.09.012},
url = {https://www.sciencedirect.com/science/article/pii/S0003491610001752},
author = {Ulrich Schollwöck},
}

@article{Bejan2024circuits,
  title = {Dynamical Magic Transitions in Monitored {C}lifford+${T}$ Circuits},
  author = {Bejan, Mircea and McLauchlan, Campbell and B\'eri, Benjamin},
  journal = {PRX Quantum},
  volume = {5},
  issue = {3},
  pages = {030332},
  numpages = {30},
  year = {2024},
  month = {Aug},
  publisher = {American Physical Society},
  doi = {10.1103/PRXQuantum.5.030332},
  url = {https://link.aps.org/doi/10.1103/PRXQuantum.5.030332}
}

@article{Bravyi2019simulationofquantum,
  doi = {10.22331/q-2019-09-02-181},
  url = {https://doi.org/10.22331/q-2019-09-02-181},
  title = {Simulation of quantum circuits by low-rank stabilizer decompositions},
  author = {Bravyi, Sergey and Browne, Dan and Calpin, Padraic and Campbell, Earl and Gosset, David and Howard, Mark},
  journal = {{Quantum}},
  issn = {2521-327X},
  publisher = {{Verein zur F{\"{o}}rderung des Open Access Publizierens in den Quantenwissenschaften}},
  volume = {3},
  pages = {181},
  month = sep,
  year = {2019}
}

@misc{hoshino2025stabilizerrenyientropyconformal,
      title={Stabilizer {R}\'enyi Entropy and Conformal Field Theory}, 
      author={Masahiro Hoshino and Masaki Oshikawa and Yuto Ashida},
      year={2025},
      eprint={2503.13599},
      archivePrefix={arXiv},
      primaryClass={quant-ph},
      url={https://arxiv.org/abs/2503.13599}, 
}

@misc{hoshino2025stabilizerrenyientropyencodes,
      title={Stabilizer {R}\'{e}nyi Entropy Encodes Fusion Rules of Topological Defects and Boundaries}, 
      author={Masahiro Hoshino and Yuto Ashida},
      year={2025},
      eprint={2507.10656},
      archivePrefix={arXiv},
      primaryClass={quant-ph},
      url={https://arxiv.org/abs/2507.10656}, 
}

@article{White_2021,
  title = {Conformal field theories are magical},
  author = {White, Christopher David and Cao, ChunJun and Swingle, Brian},
  journal = {Phys. Rev. B},
  volume = {103},
  issue = {7},
  pages = {075145},
  numpages = {18},
  year = {2021},
  month = {Feb},
  publisher = {American Physical Society},
  doi = {10.1103/PhysRevB.103.075145},
  url = {https://link.aps.org/doi/10.1103/PhysRevB.103.075145}
}

@Article{tarabunga2025efficientmutualmagicmagic,
	title={{Efficient mutual magic and magic capacity with matrix product states}},
	author={Poetri Sonya Tarabunga and Tobias Haug},
	journal={SciPost Phys.},
	volume={19},
	pages={085},
	year={2025},
	publisher={SciPost},
	doi={10.21468/SciPostPhys.19.4.085},
	url={https://scipost.org/10.21468/SciPostPhys.19.4.085},
}

@article{jasser2025stabilizerentropyentanglementcomplexity,
  title = {Stabilizer entropy and entanglement complexity in the Sachdev-Ye-Kitaev model},
  author = {Jasser, Barbara and Odavi\ifmmode \acute{c}\else \'{c}\fi{}, Jovan and Hamma, Alioscia},
  journal = {Phys. Rev. B},
  volume = {112},
  issue = {17},
  pages = {174204},
  numpages = {16},
  year = {2025},
  month = {Nov},
  publisher = {American Physical Society},
  doi = {10.1103/rz86-47h3},
  url = {https://link.aps.org/doi/10.1103/rz86-47h3}
}

@Article{bera2025nonstabilizernesssachdevyekitaevmodel,
	title={{Non-stabilizerness of Sachdev-Ye-Kitaev model}},
	author={Surajit Bera and Marco Schirò},
	journal={SciPost Phys.},
	volume={19},
	pages={159},
	year={2025},
	publisher={SciPost},
	doi={10.21468/SciPostPhys.19.6.159},
	url={https://scipost.org/10.21468/SciPostPhys.19.6.159},
}

@article{odavić2025stabilizerentropynonintegrablequantum,
  title = {Stabilizer entropy in nonintegrable quantum evolutions},
  author = {Odavi\ifmmode \acute{c}\else \'{c}\fi{}, J. and Viscardi, M. and Hamma, A.},
  journal = {Phys. Rev. B},
  volume = {112},
  issue = {10},
  pages = {104301},
  numpages = {14},
  year = {2025},
  month = {Sep},
  publisher = {American Physical Society},
  doi = {10.1103/y9r6-dx7p},
  url = {https://link.aps.org/doi/10.1103/y9r6-dx7p}
}

@misc{rajabpour2025stabilizershannonrenyiequivalenceexact,
      title={Stabilizer-Shannon Renyi Equivalence: Exact Results for Quantum Critical Chains}, 
      author={M. A. Rajabpour},
      year={2025},
      eprint={2509.10700},
      archivePrefix={arXiv},
      primaryClass={quant-ph},
      url={https://arxiv.org/abs/2509.10700}, 
}

@article{dóra2024momentumspacemagictransverse,
  title = {Momentum space nonstabilizerness for the transverse field quantum Ising model},
  author = {D\'ora, Bal\'azs and Moca, C\u{a}t\u{a}lin Pa\c{s}cu},
  journal = {Phys. Rev. B},
  volume = {112},
  issue = {12},
  pages = {125427},
  numpages = {8},
  year = {2025},
  month = {Sep},
  publisher = {American Physical Society},
  doi = {10.1103/mx8t-l4hf},
  url = {https://link.aps.org/doi/10.1103/mx8t-l4hf}
}

@misc{collura2025quantummagicfermionicgaussian,
      title={The non-stabilizerness of fermionic Gaussian states}, 
      author={Mario Collura and Jacopo De Nardis and Vincenzo Alba and Guglielmo Lami},
      year={2025},
      eprint={2412.05367},
      archivePrefix={arXiv},
      primaryClass={quant-ph},
      url={https://arxiv.org/abs/2412.05367}, 
}

@misc{moca2025quantumwalk,
      title={Non-stabilizerness generation in a multi-particle quantum walk}, 
      author={Cătălin Paşcu Moca and Doru Sticlet and Balázs Dóra and Angelo Valli and Dominik Szombathy and Gergely Zaránd},
      year={2025},
      eprint={2504.19750},
      archivePrefix={arXiv},
      primaryClass={quant-ph},
      url={https://arxiv.org/abs/2504.19750}, 
}

@article{Zauner2018VUMPS,
  title = {Variational optimization algorithms for uniform matrix product states},
  author = {Zauner-Stauber, V. and Vanderstraeten, L. and Fishman, M. T. and Verstraete, F. and Haegeman, J.},
  journal = {Phys. Rev. B},
  volume = {97},
  issue = {4},
  pages = {045145},
  numpages = {31},
  year = {2018},
  month = {Jan},
  publisher = {American Physical Society},
  doi = {10.1103/PhysRevB.97.045145},
  url = {https://link.aps.org/doi/10.1103/PhysRevB.97.045145}
}

@article{Pollmann2008,
  title = {Theory of Finite-Entanglement Scaling at One-Dimensional Quantum Critical Points},
  author = {Pollmann, Frank and Mukerjee, Subroto and Turner, Ari M. and Moore, Joel E.},
  journal = {Phys. Rev. Lett.},
  volume = {102},
  issue = {25},
  pages = {255701},
  numpages = {4},
  year = {2009},
  month = {Jun},
  publisher = {American Physical Society},
  doi = {10.1103/PhysRevLett.102.255701},
  url = {https://link.aps.org/doi/10.1103/PhysRevLett.102.255701}
}

@article{Verresen2018Topology,
  title = {Topology and Edge Modes in Quantum Critical Chains},
  author = {Verresen, Ruben and Jones, Nick G. and Pollmann, Frank},
  journal = {Phys. Rev. Lett.},
  volume = {120},
  issue = {5},
  pages = {057001},
  numpages = {5},
  year = {2018},
  month = {Jan},
  publisher = {American Physical Society},
  doi = {10.1103/PhysRevLett.120.057001},
  url = {https://link.aps.org/doi/10.1103/PhysRevLett.120.057001}
}

@article{Jones2019Topology,
author={Jones, N. G.
and Verresen, R.},
title={Asymptotic Correlations in Gapped and Critical Topological Phases of 1{D} Quantum Systems},
journal={Journal of Statistical Physics},
year={2019},
month={Jun},
day={01},
volume={175},
number={6},
pages={1164-1213},
issn={1572-9613},
doi={10.1007/s10955-019-02257-9},
url={https://doi.org/10.1007/s10955-019-02257-9}
}

@article{Altland1997SymmetryClass,
  title = {Nonstandard symmetry classes in mesoscopic normal-superconducting hybrid structures},
  author = {Altland, Alexander and Zirnbauer, Martin R.},
  journal = {Phys. Rev. B},
  volume = {55},
  issue = {2},
  pages = {1142--1161},
  numpages = {0},
  year = {1997},
  month = {Jan},
  publisher = {American Physical Society},
  doi = {10.1103/PhysRevB.55.1142},
  url = {https://link.aps.org/doi/10.1103/PhysRevB.55.1142}
}

@article{Smacchia2011statmechclusterIsing,
  title = {Statistical mechanics of the cluster Ising model},
  author = {Smacchia, Pietro and Amico, Luigi and Facchi, Paolo and Fazio, Rosario and Florio, Giuseppe and Pascazio, Saverio and Vedral, Vlatko},
  journal = {Phys. Rev. A},
  volume = {84},
  issue = {2},
  pages = {022304},
  numpages = {12},
  year = {2011},
  month = {Aug},
  publisher = {American Physical Society},
  doi = {10.1103/PhysRevA.84.022304},
  url = {https://link.aps.org/doi/10.1103/PhysRevA.84.022304}
}

@article{Leone2024monotones,
  title = {Stabilizer entropies are monotones for magic-state resource theory},
  author = {Leone, Lorenzo and Bittel, Lennart},
  journal = {Phys. Rev. A},
  volume = {110},
  issue = {4},
  pages = {L040403},
  numpages = {6},
  year = {2024},
  month = {Oct},
  publisher = {American Physical Society},
  doi = {10.1103/PhysRevA.110.L040403},
  url = {https://link.aps.org/doi/10.1103/PhysRevA.110.L040403}
}

@article{Kitaev2006,
  title = {Topological Entanglement Entropy},
  author = {Kitaev, Alexei and Preskill, John},
  journal = {Phys. Rev. Lett.},
  volume = {96},
  issue = {11},
  pages = {110404},
  numpages = {4},
  year = {2006},
  month = {Mar},
  publisher = {American Physical Society},
  doi = {10.1103/PhysRevLett.96.110404},
  url = {https://link.aps.org/doi/10.1103/PhysRevLett.96.110404}
}

@article{Levin2006,
  title = {Detecting Topological Order in a Ground State Wave Function},
  author = {Levin, Michael and Wen, Xiao-Gang},
  journal = {Phys. Rev. Lett.},
  volume = {96},
  issue = {11},
  pages = {110405},
  numpages = {4},
  year = {2006},
  month = {Mar},
  publisher = {American Physical Society},
  doi = {10.1103/PhysRevLett.96.110405},
  url = {https://link.aps.org/doi/10.1103/PhysRevLett.96.110405}
}

@article{Pollmann2012,
  title = {Symmetry protection of topological phases in one-dimensional quantum spin systems},
  author = {Pollmann, Frank and Berg, Erez and Turner, Ari M. and Oshikawa, Masaki},
  journal = {Phys. Rev. B},
  volume = {85},
  issue = {7},
  pages = {075125},
  numpages = {9},
  year = {2012},
  month = {Feb},
  publisher = {American Physical Society},
  doi = {10.1103/PhysRevB.85.075125},
  url = {https://link.aps.org/doi/10.1103/PhysRevB.85.075125}
}

@article{Li2008,
  title = {Entanglement Spectrum as a Generalization of Entanglement Entropy: Identification of Topological Order in Non-Abelian Fractional Quantum Hall Effect States},
  author = {Li, Hui and Haldane, F. D. M.},
  journal = {Phys. Rev. Lett.},
  volume = {101},
  issue = {1},
  pages = {010504},
  numpages = {4},
  year = {2008},
  month = {Jul},
  publisher = {American Physical Society},
  doi = {10.1103/PhysRevLett.101.010504},
  url = {https://link.aps.org/doi/10.1103/PhysRevLett.101.010504}
}

@article{Bardarson2012,
  title = {Unbounded Growth of Entanglement in Models of Many-Body Localization},
  author = {Bardarson, Jens H. and Pollmann, Frank and Moore, Joel E.},
  journal = {Phys. Rev. Lett.},
  volume = {109},
  issue = {1},
  pages = {017202},
  numpages = {5},
  year = {2012},
  month = {Jul},
  publisher = {American Physical Society},
  doi = {10.1103/PhysRevLett.109.017202},
  url = {https://link.aps.org/doi/10.1103/PhysRevLett.109.017202}
}

@article{Laflorencie2016,
title = {Quantum entanglement in condensed matter systems},
journal = {Physics Reports},
volume = {646},
pages = {1-59},
year = {2016},
note = {},
issn = {0370-1573},
doi = {https://doi.org/10.1016/j.physrep.2016.06.008},
url = {https://www.sciencedirect.com/science/article/pii/S0370157316301582},
author = {Nicolas Laflorencie}
}

@article{Serbyn2013,
  title = {Universal Slow Growth of Entanglement in Interacting Strongly Disordered Systems},
  author = {Serbyn, Maksym and Papi\ifmmode \acute{c}\else \'{c}\fi{}, Z. and Abanin, Dmitry A.},
  journal = {Phys. Rev. Lett.},
  volume = {110},
  issue = {26},
  pages = {260601},
  numpages = {5},
  year = {2013},
  month = {Jun},
  publisher = {American Physical Society},
  doi = {10.1103/PhysRevLett.110.260601},
  url = {https://link.aps.org/doi/10.1103/PhysRevLett.110.260601}
}

@article{lukin2019,
   title={Probing entanglement in a many-body–localized system},
   volume={364},
   ISSN={1095-9203},
   url={http://dx.doi.org/10.1126/science.aau0818},
   DOI={10.1126/science.aau0818},
   number={6437},
   journal={Science},
   publisher={American Association for the Advancement of Science (AAAS)},
   author={Lukin, Alexander and Rispoli, Matthew and Schittko, Robert and Tai, M. Eric and Kaufman, Adam M. and Choi, Soonwon and Khemani, Vedika and Léonard, Julian and Greiner, Markus},
   year={2019},
   month=apr, pages={256–260} }

@article{CiracRMP,
  title = {Matrix product states and projected entangled pair states: Concepts, symmetries, theorems},
  author = {Cirac, J. Ignacio and P\'erez-Garc\'{\i}a, David and Schuch, Norbert and Verstraete, Frank},
  journal = {Rev. Mod. Phys.},
  volume = {93},
  issue = {4},
  pages = {045003},
  numpages = {65},
  year = {2021},
  month = {Dec},
  publisher = {American Physical Society},
  doi = {10.1103/RevModPhys.93.045003},
  url = {https://link.aps.org/doi/10.1103/RevModPhys.93.045003}
}

@article{WhiteDMRG,
  title = {Density matrix formulation for quantum renormalization groups},
  author = {White, Steven R.},
  journal = {Phys. Rev. Lett.},
  volume = {69},
  issue = {19},
  pages = {2863--2866},
  numpages = {0},
  year = {1992},
  month = {Nov},
  publisher = {American Physical Society},
  doi = {10.1103/PhysRevLett.69.2863},
  url = {https://link.aps.org/doi/10.1103/PhysRevLett.69.2863}
}

@article{Eastin2009,
  title = {Restrictions on Transversal Encoded Quantum Gate Sets},
  author = {Eastin, Bryan and Knill, Emanuel},
  journal = {Phys. Rev. Lett.},
  volume = {102},
  issue = {11},
  pages = {110502},
  numpages = {4},
  year = {2009},
  month = {Mar},
  publisher = {American Physical Society},
  doi = {10.1103/PhysRevLett.102.110502},
  url = {https://link.aps.org/doi/10.1103/PhysRevLett.102.110502}
}

@misc{gottesman1998,
      title={The Heisenberg Representation of Quantum Computers}, 
      author={Daniel Gottesman},
      year={1998},
      eprint={quant-ph/9807006},
      archivePrefix={arXiv},
      primaryClass={quant-ph}
}

@misc{knill2004,
      title={Fault-Tolerant Postselected Quantum Computation: Schemes}, 
      author={E. Knill},
      year={2004},
      eprint={quant-ph/0402171},
      archivePrefix={arXiv},
      primaryClass={quant-ph}
}

@article{Bravyi2005,
  title = {Universal quantum computation with ideal Clifford gates and noisy ancillas},
  author = {Bravyi, Sergey and Kitaev, Alexei},
  journal = {Phys. Rev. A},
  volume = {71},
  issue = {2},
  pages = {022316},
  numpages = {14},
  year = {2005},
  month = {Feb},
  publisher = {American Physical Society},
  doi = {10.1103/PhysRevA.71.022316},
  url = {https://link.aps.org/doi/10.1103/PhysRevA.71.022316}
}

@article{Campbell2010,
  title = {Bound States for Magic State Distillation in Fault-Tolerant Quantum Computation},
  author = {Campbell, Earl T. and Browne, Dan E.},
  journal = {Phys. Rev. Lett.},
  volume = {104},
  issue = {3},
  pages = {030503},
  numpages = {4},
  year = {2010},
  month = {Jan},
  publisher = {American Physical Society},
  doi = {10.1103/PhysRevLett.104.030503},
  url = {https://link.aps.org/doi/10.1103/PhysRevLett.104.030503}
}

@Article{ShiyuTgate2020,
	title={{Single T gate in a Clifford circuit drives transition to universal entanglement spectrum statistics}},
	author={Shiyu Zhou and Zhi-Cheng Yang and Alioscia Hamma and Claudio Chamon},
	journal={SciPost Phys.},
	volume={9},
	pages={087},
	year={2020},
	publisher={SciPost},
	doi={10.21468/SciPostPhys.9.6.087},
	url={https://scipost.org/10.21468/SciPostPhys.9.6.087},
}

@article{Liu2022,
  title = {Many-Body Quantum Magic},
  author = {Liu, Zi-Wen and Winter, Andreas},
  journal = {PRX Quantum},
  volume = {3},
  issue = {2},
  pages = {020333},
  numpages = {18},
  year = {2022},
  month = {May},
  publisher = {American Physical Society},
  doi = {10.1103/PRXQuantum.3.020333},
  url = {https://link.aps.org/doi/10.1103/PRXQuantum.3.020333}
}

@article{goto2021chaos,
  title = {Probing chaos by magic monotones},
  author = {Goto, Kanato and Nosaka, Tomoki and Nozaki, Masahiro},
  journal = {Phys. Rev. D},
  volume = {106},
  issue = {12},
  pages = {126009},
  numpages = {26},
  year = {2022},
  month = {Dec},
  publisher = {American Physical Society},
  doi = {10.1103/PhysRevD.106.126009},
  url = {https://link.aps.org/doi/10.1103/PhysRevD.106.126009}
}

@article{Turkeshi2023,
  title = {Measuring nonstabilizerness via multifractal flatness},
  author = {Turkeshi, Xhek and Schir\`o, Marco and Sierant, Piotr},
  journal = {Phys. Rev. A},
  volume = {108},
  issue = {4},
  pages = {042408},
  numpages = {8},
  year = {2023},
  month = {Oct},
  publisher = {American Physical Society},
  doi = {10.1103/PhysRevA.108.042408},
  url = {https://link.aps.org/doi/10.1103/PhysRevA.108.042408}
}

@book{Nielsen_Chuang_2010, place={Cambridge}, title={Quantum Computation and Quantum Information: 10th Anniversary Edition}, publisher={Cambridge University Press}, author={Nielsen, Michael A. and Chuang, Isaac L.}, year={2010}}

@article{frau2024nonstabilizerness,
  title = {Nonstabilizerness versus entanglement in matrix product states},
  author = {Frau, M. and Tarabunga, P. S. and Collura, M. and Dalmonte, M. and Tirrito, E.},
  journal = {Phys. Rev. B},
  volume = {110},
  issue = {4},
  pages = {045101},
  numpages = {13},
  year = {2024},
  month = {Jul},
  publisher = {American Physical Society},
  doi = {10.1103/PhysRevB.110.045101},
  url = {https://link.aps.org/doi/10.1103/PhysRevB.110.045101}
}

@article{tarabunga2024critical,
  doi = {10.22331/q-2024-07-17-1413},
  url = {https://doi.org/10.22331/q-2024-07-17-1413},
  title = {Critical behaviors of non-stabilizerness in quantum spin chains},
  author = {Tarabunga, Poetri Sonya},
  journal = {{Quantum}},
  issn = {2521-327X},
  publisher = {{Verein zur F{\"{o}}rderung des Open Access Publizierens in den Quantenwissenschaften}},
  volume = {8},
  pages = {1413},
  month = jul,
  year = {2024}
}

@article{Tarabunga2024RK,
   title={Magic in generalized {R}okhsar-{K}ivelson wavefunctions},
   volume={8},
   ISSN={2521-327X},
   url={http://dx.doi.org/10.22331/q-2024-05-14-1347},
   DOI={10.22331/q-2024-05-14-1347},
   journal={Quantum},
   publisher={Verein zur Forderung des Open Access Publizierens in den Quantenwissenschaften},
   author={Tarabunga, Poetri Sonya and Castelnovo, Claudio},
   year={2024},
   month=may, pages={1347} }

@article{Calabrese2004,
doi = {10.1088/1742-5468/2004/06/P06002},
url = {https://doi.org/10.1088/1742-5468/2004/06/P06002},
year = {2004},
month = {jun},
publisher = {},
volume = {2004},
number = {06},
pages = {P06002},
author = {Pasquale Calabrese and John Cardy},
title = {Entanglement entropy and quantum field theory},
journal = {Journal of Statistical Mechanics: Theory and Experiment}
}

@article{Tarabunga2023,
  title = {Many-Body Magic Via Pauli-Markov Chains---From Criticality to Gauge Theories},
  author = {Tarabunga, Poetri Sonya and Tirrito, Emanuele and Chanda, Titas and Dalmonte, Marcello},
  journal = {PRX Quantum},
  volume = {4},
  issue = {4},
  pages = {040317},
  numpages = {19},
  year = {2023},
  month = {Oct},
  publisher = {American Physical Society},
  doi = {10.1103/PRXQuantum.4.040317},
  url = {https://link.aps.org/doi/10.1103/PRXQuantum.4.040317}
}

@article{Leone2022,
  title = {Stabilizer {R}\'enyi Entropy},
  author = {Leone, Lorenzo and Oliviero, Salvatore F. E. and Hamma, Alioscia},
  journal = {Phys. Rev. Lett.},
  volume = {128},
  issue = {5},
  pages = {050402},
  numpages = {5},
  year = {2022},
  month = {Feb},
  publisher = {American Physical Society},
  doi = {10.1103/PhysRevLett.128.050402},
  url = {https://link.aps.org/doi/10.1103/PhysRevLett.128.050402}
}

@article{Haug2025probingquantum,
  doi = {10.22331/q-2025-07-21-1801},
  url = {https://doi.org/10.22331/q-2025-07-21-1801},
  title = {Probing quantum complexity via universal saturation of stabilizer entropies},
  author = {Haug, Tobias and Aolita, Leandro and Kim, M.S.},
  journal = {{Quantum}},
  issn = {2521-327X},
  publisher = {{Verein zur F{\"{o}}rderung des Open Access Publizierens in den Quantenwissenschaften}},
  volume = {9},
  pages = {1801},
  month = jul,
  year = {2025}
}

@misc{liu2025SRETIMPS,
      title={Stabilizer {R}\'enyi Entropy for Translation-Invariant Matrix Product States}, 
      author={Lei-Yi-Nan Liu and Su Yi and Jian Cui},
      year={2025},
      eprint={2508.03534},
      archivePrefix={arXiv},
      primaryClass={quant-ph},
      url={https://arxiv.org/abs/2508.03534}, 
}

@article{catalano2025SREtopphases,
  title = {Resource complexity of symmetry-protected topological phases},
  author = {Catalano, Alberto Giuseppe and Ko\ifmmode \check{z}\else \v{z}\fi{}i\ifmmode \acute{c}\else \'{c}\fi{}, Sven Benjamin and Torre, Gianpaolo and Ciaramelletti, Carola and Paganelli, Simone and Franchini, Fabio and Giampaolo, Salvatore Marco},
  journal = {Phys. Rev. B},
  volume = {113},
  issue = {15},
  pages = {155126},
  numpages = {8},
  year = {2026},
  month = {Apr},
  publisher = {American Physical Society},
  doi = {10.1103/t4fm-t7tc},
  url = {https://link.aps.org/doi/10.1103/t4fm-t7tc}
}

@misc{haug2025efficientwitnessingtestingmagic,
      title={Efficient witnessing and testing of magic in mixed quantum states}, 
      author={Tobias Haug and Poetri Sonya Tarabunga},
      year={2025},
      eprint={2504.18098},
      archivePrefix={arXiv},
      primaryClass={quant-ph},
      url={https://arxiv.org/abs/2504.18098}, 
}

@misc{moca2025nonstabilizernessdiagnosticcriticalityexceptional,
      title={Non-stabilizerness as a Diagnostic of Criticality and Exceptional Points in Non-Hermitian Spin Chains}, 
      author={Cătălin Paşcu Moca and Doru Sticlet and Balázs Dóra},
      year={2025},
      eprint={2510.17248},
      archivePrefix={arXiv},
      primaryClass={quant-ph},
      url={https://arxiv.org/abs/2510.17248}, 
}

@article{Ding2025singularites,
  title = {Evaluating Many-Body Stabilizer {R}\'enyi Entropy by Sampling Reduced {P}auli Strings: Singularities, Volume Law, and Nonlocal Magic},
  author = {Ding, Yi-Ming and Wang, Zhe and Yan, Zheng},
  journal = {PRX Quantum},
  volume = {6},
  issue = {3},
  pages = {030328},
  numpages = {15},
  year = {2025},
  month = {Aug},
  publisher = {American Physical Society},
  doi = {10.1103/pyzr-jmvw},
  url = {https://link.aps.org/doi/10.1103/pyzr-jmvw}
}

@article{cao2025gravitationalbackreactionmagical,
  title = {Gravitational Backreaction is Magical},
  author = {Cao, ChunJun and Cheng, Gong and Hamma, Alioscia and Leone, Lorenzo and Munizzi, William and Oliviero, Savatore F.E.},
  journal = {PRX Quantum},
  volume = {6},
  issue = {4},
  pages = {040375},
  numpages = {39},
  year = {2025},
  month = {Dec},
  publisher = {American Physical Society},
  doi = {10.1103/z3vr-w5c5},
  url = {https://link.aps.org/doi/10.1103/z3vr-w5c5}
}

@article{Passarelli2024permutationallyinvariant,
  title = {Nonstabilizerness of permutationally invariant systems},
  author = {Passarelli, G. and Fazio, R. and Lucignano, P.},
  journal = {Phys. Rev. A},
  volume = {110},
  issue = {2},
  pages = {022436},
  numpages = {9},
  year = {2024},
  month = {Aug},
  publisher = {American Physical Society},
  doi = {10.1103/PhysRevA.110.022436},
  url = {https://link.aps.org/doi/10.1103/PhysRevA.110.022436}
}

@article{Melko2010montecarloMI,
  title = {Finite-size scaling of mutual information in Monte Carlo simulations: Application to the spin-$\frac{1}{2}$ {XXZ} model},
  author = {Melko, Roger G. and Kallin, Ann B. and Hastings, Matthew B.},
  journal = {Phys. Rev. B},
  volume = {82},
  issue = {10},
  pages = {100409},
  numpages = {4},
  year = {2010},
  month = {Sep},
  publisher = {American Physical Society},
  doi = {10.1103/PhysRevB.82.100409},
  url = {https://link.aps.org/doi/10.1103/PhysRevB.82.100409}
}

@article{Wilms_2011Mutualinfo,
doi = {10.1088/1742-5468/2011/10/P10011},
url = {https://doi.org/10.1088/1742-5468/2011/10/P10011},
year = {2011},
month = {oct},
publisher = {},
volume = {2011},
number = {10},
pages = {P10011},
author = {Wilms, Johannes and Troyer, Matthias and Verstraete, Frank},
title = {Mutual information in classical spin models},
journal = {Journal of Statistical Mechanics: Theory and Experiment}
}

@article{Wolf2008MI,
  title = {Area Laws in Quantum Systems: Mutual Information and Correlations},
  author = {Wolf, Michael M. and Verstraete, Frank and Hastings, Matthew B. and Cirac, J. Ignacio},
  journal = {Phys. Rev. Lett.},
  volume = {100},
  issue = {7},
  pages = {070502},
  numpages = {4},
  year = {2008},
  month = {Feb},
  publisher = {American Physical Society},
  doi = {10.1103/PhysRevLett.100.070502},
  url = {https://link.aps.org/doi/10.1103/PhysRevLett.100.070502}
}

@article{Kudler2023,
  title = {R\'enyi Mutual Information in Quantum Field Theory},
  author = {Kudler-Flam, Jonah},
  journal = {Phys. Rev. Lett.},
  volume = {130},
  issue = {2},
  pages = {021603},
  numpages = {6},
  year = {2023},
  month = {Jan},
  publisher = {American Physical Society},
  doi = {10.1103/PhysRevLett.130.021603},
  url = {https://link.aps.org/doi/10.1103/PhysRevLett.130.021603}
}

@Article{Fliss2021,
author={Fliss, Jackson R.},
title={Knots, links, and long-range magic},
journal={Journal of High Energy Physics},
year={2021},
month={Apr},
day={09},
volume={2021},
number={4},
pages={90},
issn={1029-8479},
doi={10.1007/JHEP04(2021)090},
url={https://doi.org/10.1007/JHEP04(2021)090}
}

@article{Bao2022,
  title = {Magic state distillation from entangled states},
  author = {Bao, Ning and Cao, ChunJun and Su, Vincent Paul},
  journal = {Phys. Rev. A},
  volume = {105},
  issue = {2},
  pages = {022602},
  numpages = {13},
  year = {2022},
  month = {Feb},
  publisher = {American Physical Society},
  doi = {10.1103/PhysRevA.105.022602},
  url = {https://link.aps.org/doi/10.1103/PhysRevA.105.022602}
}

@article{Dowling2025Heisenberg,
  title = {Magic Resources of the {H}eisenberg Picture},
  author = {Dowling, Neil and Kos, Pavel and Turkeshi, Xhek},
  journal = {Phys. Rev. Lett.},
  volume = {135},
  issue = {5},
  pages = {050401},
  numpages = {10},
  year = {2025},
  month = {Jul},
  publisher = {American Physical Society},
  doi = {10.1103/p7xt-s9nz},
  url = {https://link.aps.org/doi/10.1103/p7xt-s9nz}
}

@article{sierant2025fermionicmagicresourcesquantum,
  title = {Fermionic Magic Resources of Quantum Many-Body Systems},
  author = {Sierant, Piotr and Stornati, Paolo and Turkeshi, Xhek},
  journal = {PRX Quantum},
  volume = {7},
  issue = {1},
  pages = {010302},
  numpages = {41},
  year = {2026},
  month = {Jan},
  publisher = {American Physical Society},
  doi = {10.1103/3yx4-1j27},
  url = {https://link.aps.org/doi/10.1103/3yx4-1j27}
}

@misc{wang2025magictransitionmonitoredfree,
      title={Magic transition in monitored free fermion dynamics}, 
      author={Cheng Wang and Zhi-Cheng Yang and Tianci Zhou and Xiao Chen},
      year={2025},
      eprint={2507.10688},
      archivePrefix={arXiv},
      primaryClass={quant-ph},
      url={https://arxiv.org/abs/2507.10688}, 
}

@article{dowling2025bridgingentanglementmagicresources,
  title = {Bridging Entanglement and Magic Resources within Operator Space},
  author = {Dowling, Neil and Modi, Kavan and White, Gregory A. L.},
  journal = {Phys. Rev. Lett.},
  volume = {135},
  issue = {16},
  pages = {160201},
  numpages = {9},
  year = {2025},
  month = {Oct},
  publisher = {American Physical Society},
  doi = {10.1103/c7k1-xcwy},
  url = {https://link.aps.org/doi/10.1103/c7k1-xcwy}
}

@article{Tagliacozzo2008,
  title = {Scaling of entanglement support for matrix product states},
  author = {Tagliacozzo, L. and de Oliveira, Thiago. R. and Iblisdir, S. and Latorre, J. I.},
  journal = {Phys. Rev. B},
  volume = {78},
  issue = {2},
  pages = {024410},
  numpages = {14},
  year = {2008},
  month = {Jul},
  publisher = {American Physical Society},
  doi = {10.1103/PhysRevB.78.024410},
  url = {https://link.aps.org/doi/10.1103/PhysRevB.78.024410}
}

@misc{sarkis2025magichybridbosonfermionsystems,
      title={Magic for Hybrid Boson-Fermion Systems: A {G}rassmann Phase-Space Approach}, 
      author={Matthieu Sarkis and Pablo Martinez-Azcona and Alexandre Tkatchenko},
      year={2025},
      eprint={2509.05264},
      archivePrefix={arXiv},
      primaryClass={quant-ph},
      url={https://arxiv.org/abs/2509.05264}, 
}

@article{crew2025magicentropyhybridspinboson,
doi = {10.1088/1361-6633/ae413c},
url = {https://doi.org/10.1088/1361-6633/ae413c},
year = {2026},
month = {feb},
publisher = {IOP Publishing},
volume = {89},
number = {2},
pages = {027602},
author = {Crew, Samuel and Li, Ying-Lin and Li, Heng-Hsi and Chang, Po-Yao},
title = {Magic entropy in hybrid spin-boson systems},
journal = {Reports on Progress in Physics},
}

@article{zhang2025stabilizerrenyientropytransition,
  title = {Stabilizer R\'enyi Entropy and Its Transition in the Coupled Sachdev-Ye-Kitaev Model},
  author = {Zhang, Pengfei and Zhou, Shuyan and Sun, Ning},
  journal = {Phys. Rev. Lett.},
  volume = {136},
  issue = {8},
  pages = {080201},
  numpages = {7},
  year = {2026},
  month = {Feb},
  publisher = {American Physical Society},
  doi = {10.1103/5c15-4g5n},
  url = {https://link.aps.org/doi/10.1103/5c15-4g5n}
}

@Article{Turkeshi2025randomqc,
author={Turkeshi, Xhek
and Tirrito, Emanuele
and Sierant, Piotr},
title={Magic spreading in random quantum circuits},
journal={Nature Communications},
year={2025},
month={Mar},
day={15},
volume={16},
number={1},
pages={2575},
issn={2041-1723},
doi={10.1038/s41467-025-57704-x},
url={https://doi.org/10.1038/s41467-025-57704-x}
}

@article{Russomanno2025SYK,
  title = {Nonstabilizerness in the unitary and monitored quantum dynamics of {XXZ}-staggered and {S}achdev-{Y}e-{K}itaev models},
  author = {Russomanno, Angelo and Passarelli, Gianluca and Rossini, Davide and Lucignano, Procolo},
  journal = {Phys. Rev. B},
  volume = {112},
  issue = {6},
  pages = {064312},
  numpages = {11},
  year = {2025},
  month = {Aug},
  publisher = {American Physical Society},
  doi = {10.1103/njgn-fksh},
  url = {https://link.aps.org/doi/10.1103/njgn-fksh}
}

@article{falcao2025magicdynamicsmanybodylocalized,
  title = {Nonstabilizerness Dynamics in Many-Body Localized Systems},
  author = {Falc\~ao, Pedro R. Nic\'acio and Sierant, Piotr and Zakrzewski, Jakub and Tirrito, Emanuele},
  journal = {Phys. Rev. Lett.},
  volume = {135},
  issue = {24},
  pages = {240404},
  numpages = {11},
  year = {2025},
  month = {Dec},
  publisher = {American Physical Society},
  doi = {10.1103/xfp5-hhs4},
  url = {https://link.aps.org/doi/10.1103/xfp5-hhs4}
}

@article{hernándezyanes2025nonstabilizernessquantumenhancedmetrologicalprotocols,
  title = {Nonstabilizerness in quantum-enhanced metrological protocols},
  author = {Hern\'andez-Yanes, Tanaus\'u and Sierant, Piotr and Zakrzewski, Jakub and P\l{}odzie\ifmmode \acute{n}\else \'{n}\fi{}, Marcin},
  journal = {Phys. Rev. A},
  volume = {113},
  issue = {1},
  pages = {012416},
  numpages = {16},
  year = {2026},
  month = {Jan},
  publisher = {American Physical Society},
  doi = {10.1103/tmf9-fyc2},
  url = {https://link.aps.org/doi/10.1103/tmf9-fyc2}
}

@article{tirrito2025anticoncentrationnonstabilizernessspreadingergodic,
  title = {Anticoncentration and Nonstabilizerness Spreading under Ergodic Quantum Dynamics},
  author = {Tirrito, Emanuele and Turkeshi, Xhek and Sierant, Piotr},
  journal = {Phys. Rev. Lett.},
  volume = {135},
  issue = {22},
  pages = {220401},
  numpages = {9},
  year = {2025},
  month = {Nov},
  publisher = {American Physical Society},
  doi = {10.1103/1jzy-sk9r},
  url = {https://link.aps.org/doi/10.1103/1jzy-sk9r}
}

@article{Haag2023MI,
  title = {Typical Correlation Length of Sequentially Generated Tensor Network States},
  author = {Haag, Daniel and Baccari, Flavio and Styliaris, Georgios},
  journal = {PRX Quantum},
  volume = {4},
  issue = {3},
  pages = {030330},
  numpages = {43},
  year = {2023},
  month = {Aug},
  publisher = {American Physical Society},
  doi = {10.1103/PRXQuantum.4.030330},
  url = {https://link.aps.org/doi/10.1103/PRXQuantum.4.030330}
}

@article{Rubboli2024mixedstate,
  doi = {10.22331/q-2024-10-04-1492},
  url = {https://doi.org/10.22331/q-2024-10-04-1492},
  title = {Mixed-state additivity properties of magic monotones based on quantum relative entropies for single-qubit states and beyond},
  author = {Rubboli, Roberto and Takagi, Ryuji and Tomamichel, Marco},
  journal = {{Quantum}},
  issn = {2521-327X},
  publisher = {{Verein zur F{\"{o}}rderung des Open Access Publizierens in den Quantenwissenschaften}},
  volume = {8},
  pages = {1492},
  month = oct,
  year = {2024}
}

@article{López_2024XXZ,
doi = {10.1088/1751-8121/ad85b0},
url = {https://doi.org/10.1088/1751-8121/ad85b0},
year = {2024},
month = {nov},
publisher = {IOP Publishing},
volume = {57},
number = {47},
pages = {475301},
author = {Montañà López, Jordi Arnau and Kos, Pavel},
title = {Exact solution of long-range stabilizer {R}ényi entropy in the dual-unitary {XXZ} model},
journal = {Journal of Physics A: Mathematical and Theoretical}
}

@misc{maity2025localspreadingstabilizerrenyi,
      title={Local spreading of stabilizer {R}\'enyi entropy in a brickwork random {C}lifford circuit}, 
      author={Somnath Maity and Ryusuke Hamazaki},
      year={2025},
      eprint={2511.07769},
      archivePrefix={arXiv},
      primaryClass={quant-ph},
      url={https://arxiv.org/abs/2511.07769}, 
}

@misc{camp2025matrixproductstateskeletonsonsagerintegrable,
      title={Matrix-product state skeletons in {O}nsager-integrable quantum chains}, 
      author={Imogen Camp and Nick G. Jones},
      year={2025},
      eprint={2511.07212},
      archivePrefix={arXiv},
      primaryClass={quant-ph},
      url={https://arxiv.org/abs/2511.07212}, 
}

@article{Hastings2010Renyiswaptrick,
  title = {Measuring {R}enyi Entanglement Entropy in Quantum Monte Carlo Simulations},
  author = {Hastings, Matthew B. and Gonz\'alez, Iv\'an and Kallin, Ann B. and Melko, Roger G.},
  journal = {Phys. Rev. Lett.},
  volume = {104},
  issue = {15},
  pages = {157201},
  numpages = {4},
  year = {2010},
  month = {Apr},
  publisher = {American Physical Society},
  doi = {10.1103/PhysRevLett.104.157201},
  url = {https://link.aps.org/doi/10.1103/PhysRevLett.104.157201}
}

@article{Zou2016MPSswaptick,
  title = {Spurious long-range entanglement and replica correlation length},
  author = {Zou, Liujun and Haah, Jeongwan},
  journal = {Phys. Rev. B},
  volume = {94},
  issue = {7},
  pages = {075151},
  numpages = {17},
  year = {2016},
  month = {Aug},
  publisher = {American Physical Society},
  doi = {10.1103/PhysRevB.94.075151},
  url = {https://link.aps.org/doi/10.1103/PhysRevB.94.075151}
}

@misc{boesl2025skeletonisometrictensornetwork,
      title={Skeleton of isometric Tensor Network States for Abelian String-Net Models}, 
      author={Julian Boesl and Yu-Jie Liu and Frank Pollmann and Michael Knap},
      year={2025},
      eprint={2511.13821},
      archivePrefix={arXiv},
      primaryClass={quant-ph},
      url={https://arxiv.org/abs/2511.13821}, 
}

@misc{bejan2025magicspreadingunitaryclifford,
      title={Magic spreading under unitary {C}lifford dynamics}, 
      author={Mircea Bejan and Pieter W. Claeys and Jiangtian Yao},
      year={2025},
      eprint={2511.21487},
      archivePrefix={arXiv},
      primaryClass={quant-ph},
      url={https://arxiv.org/abs/2511.21487}, 
}

@misc{nehra2025topologicalmagicresponsequantum,
      title={Topological magic response in quantum spin chains}, 
      author={Ritu Nehra and Poetri Sonya Tarabunga and Martina Frau and Mario Collura and Emanuele Tirrito and Marcello Dalmonte},
      year={2025},
      eprint={2512.16673},
      archivePrefix={arXiv},
      primaryClass={quant-ph},
      url={https://arxiv.org/abs/2512.16673}, 
}

@misc{data_manuscript,
  author = {Andrew Hallam and Ryan Smith and Zlatko Papi\'c},
  title = {Data repository for `Spectral signatures of nonstabilizerness and criticality in infinite matrix product states'},
  howpublished = {\url{https://doi.org/10.5518/1852}},
  year = {2026},
  note = {Data repository; DOI: 10.5518/1852},
}

\end{document}